\documentclass[traditabstract]{aa} 
\usepackage{graphicx}
\usepackage{txfonts}
\usepackage{natbib}
\usepackage{mathrsfs}

\newcommand{\vsin}{$v$~sin~$i$ }

\newcommand{\tef}{$T_{\rm eff}$ }
\newcommand{\lgg}{{\rm log}~$g$ }
\newcommand{\ms}{M$_{\odot}$}
\newcommand{\rs}{R$_{\odot}$}
\newcommand{\cd}{d$^{-1}$}

\newcommand {\logg}{\log g}

\newcommand {\eb}{KIC~3858884~}
\newcommand{\kepler}{{\it Kepler~}}
\newcommand{\kepbf}{\textbf{\textit {Kepler~}}}

\newcommand{\kms}{km~s$^{-1}$ }

\begin{document}
   \title{KIC~3858884: a hybrid $\delta$~Sct pulsator in a highly eccentric eclipsing binary.\thanks{
  Based on photometry collected by the {\it Kepler} space mission.
 }
 }
   \author{
   	C.~ Maceroni\inst{1}
	\and 
	   H.~Lehmann\inst{2}
	\and 
	R.~da Silva \inst{1}
	\and
 	  J. Montalb\'{a}n\inst{3}
   	\and 
	   C.-U.~Lee \inst{4}
   	\and 
	H.~Ak\inst{5,6}
	\and
	R.~Deshpande\inst{6,7}
	\and
  	K.~Yakut \inst{8}
	\and
	J.~Debosscher\inst{9}
	\and
	Z.~Guo\inst{10}
	\and
	S.-L.~Kim\inst{4}
	\and
	J.~W.~Lee\inst{4}
	\and
	J.~Southworth\inst{11}
	          }
   \institute{INAF--Osservatorio astronomico di Roma, via Frascati 33, I-00040 Monteporzio C., Italy\\
              \email{maceroni@oa-roma.inaf.it}
	\and 
	Th\"{u}ringer Landessternwarte Tautenburg, Sternwarte 5, D-07778 Tautenburg, Germany.
         \and
             Institut d'Astrophysique et G\'{e}ophysique Universit\'{e} de Li\`{e}ge,
		All\'{e}e du 6 A\^{o}ut, B-4000 Li\`{e}ge, Belgium. 
	\and
	Korea Astronomy and Space Science Institute, Daejeon 305-348, Korea
	\and
	Erciyes University, Science Faculty, Astronomy and Space Sci. Dept., 38039 Kayseri, Turkey
	\and
	Department of Astronomy and Astrophysics, The Pennsylvania State University, University Park, PA 16802
	\and
	Center for Exoplanets and Habitable Worlds, The Pennsylvania State University, University Park, PA 16802
	\and
	Department of Astronomy \& Space Sciences, University of Ege, 35100, \.{I}zmir, Turkey
	\and
	Instituut for Sterrenkunde, K.U.Leuven, Celestijnenlaan 200 D, B-3001, Leuven, Belgium
	\and
	Center for High Angular Resolution Astronomy and Department of Physics and Astronomy, Georgia State University,
	P.O. Box 5060, Atlanta, GA 30302-5060, USA
	\and
	Astrophysics Group, Keele University, Staffordshire, ST5 5BG, UK
             }
   \date{Received October, 18,  2013; accepted January 12, 2014}
\authorrunning{C. Maceroni et al.}

  \abstract
   {The analysis of eclipsing binaries containing non-radial pulsators allows: i) to combine two different and independent sources of information on the internal structure and evolutionary status of the components,  and ii) to study the effects of tidal forces on pulsations.  \eb is a bright \kepler target whose light curve shows deep eclipses, complex pulsation patterns with pulsation frequencies typical of $\delta$~Sct,  and a highly eccentric orbit. 
   We present the result of the analysis of \kepler photometry and of high resolution phase-resolved spectroscopy.
Spectroscopy yielded both the  radial velocity curves and, after spectral disentangling, the primary component effective temperature and  metallicity, and line-of-sight projected rotational velocities.   The \kepler light curve was  analyzed with  an iterative procedure devised to disentangle  eclipses  from pulsations which takes into account the visibility of the pulsating star during eclipses.  The search for the best set of binary parameters was performed combining the synthetic light curve models with a genetic minimization algorithm,
which yielded a robust and accurate determination of the system parameters. The binary components have very similar masses (1.88 and 1.86 \ms) and effective temperatures (6800 and 6600 K), but different radii (3.45 and 3.05 \rs).   The comparison with the theoretical models evidenced a somewhat different evolutionary status of the components and the need of introducing overshooting in the models. The pulsation analysis   indicates a hybrid nature of the pulsating (secondary) component, the corresponding high order g-modes  might  be excited  by an intrinsic mechanism or by  tidal 
forces.
 }
   
   \maketitle
   \begin{figure*}[ht!] 
  \centering
   \includegraphics[]{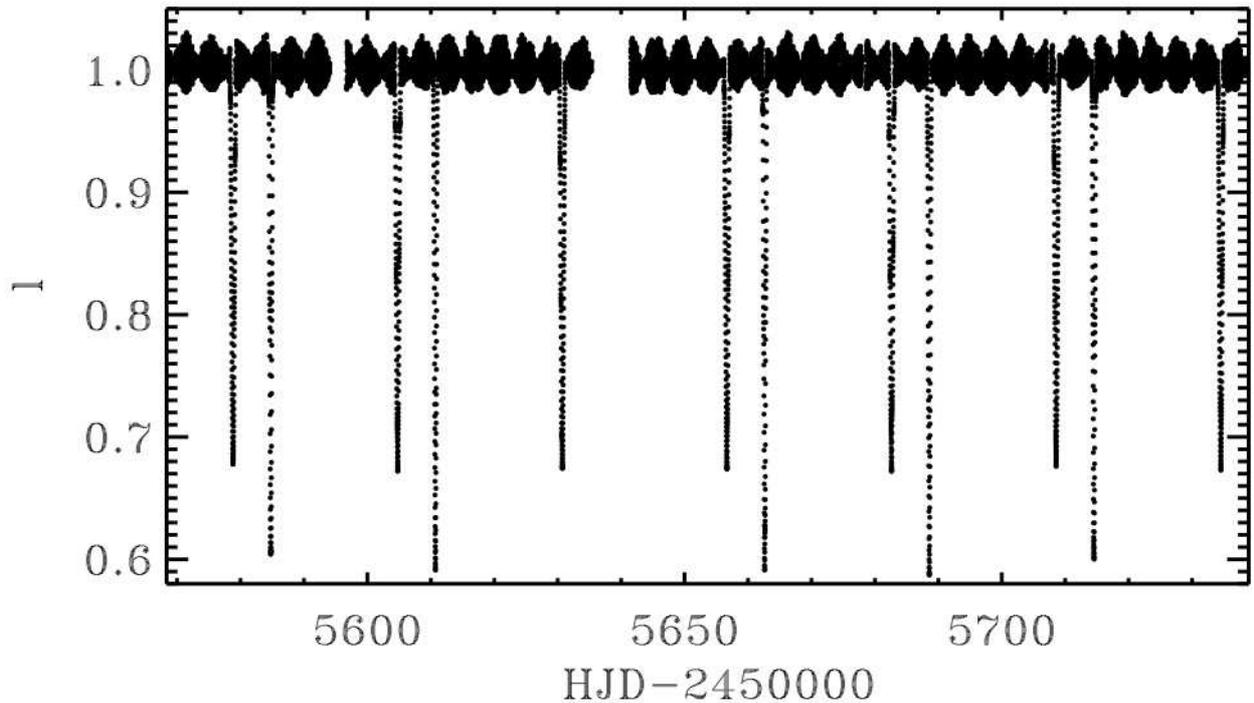}
   \caption{\eb de-trended  light-curve  of Q8 and Q9 quarters, after  rebinning   and  normalization to the quarter median.  Normalized flux,  is  denoted with $l$.}
        \label{lc}
          \end{figure*}
\section{Introduction}
 The primary science objective  of  the \kepler mission was the detection  of  earth-like planets by the transit technique,  but  the long-term almost continuous monitoring of about 160000 stars provided precious data for different fields of astrophysics, first of all for asteroseismology which  requires, as transiting planet search, accurate and continuous time series. 
An important by-product of the mission is the discovery of a large number variable stars, including  $\sim 2100$ eclipsing binaries (EBs) \citep{slawson11,prsa11,mati12}.  In some cases the binary contains a pulsating component, including  non-radial pulsators  as $\delta$ Sct, $\gamma$ Dor and Slowly Pulsating B (SPB) stars.  This providential  occurrence allows the combination of  independent information  from  two different phenomena, and their synergy in terms of  the scientific results goes well beyond  those from the single sources.
  Asteroseismology can yield  deep  insights in stellar structures, and pulsating EBs have a fundamental asset: studying non-radial oscillations in  EB components has the advantage that the masses and radii  can be independently derived,  with a pure geometrical method, by combining the information from the light and  the radial velocity  curves, and with uncertainties, in the best cases, below 1\% \citep[e.g.,][]{South05}. Moreover, additional useful constraints derive from the requirement of same chemical composition and age.
  Since the precise measurement of the mass and radius  poses  powerful constraints on  the pulsational properties,  EBs with pulsating component can provide direct tests of the modeling of complex dynamical processes occurring in stellar interiors (such as mixing length, convective overshooting, diffusion). 

The potential of pulsating EBs has been exploited in several  studies, e.g. those on HD~184884, KIC~8410637, 
 CoRoT~102918586,  KIC~11285625     \citep[][respectively] {cm09, hekker10, cm13, jonas13},  and that on a sample of \kepler binaries with red-giant components by \cite{gaulme13} .

The trade off with these advantages  is the  more complex structure and analysis of the data, as it is necessary to disentangle  the  two phenomena at the origin of the observed variability.
The problems met in this procedure  can be more or less severe, depending on the relevance of the two phenomena in modeling the light curve. Their interplay was a minor problem in the above mentioned cases since   HD 184884 \citep{cm09} and CoRoT~102918586 \citep{cm13}  show  grazing eclipses, and for KIC~8410637 (a  long period binary containing a pulsating giant)  the analysis of the typical solar like pulsations could be done in the long phase interval outside eclipses unaffected by binarity. When the effect of pulsations or eclipses cannot be treated as a perturbation, or analyzed separately, a more complex treatment is needed. In particular, when the eclipse hides from view a large part of the pulsator, its decreasing contribution to the total light shall be taken into account.
 
Variables of  $\delta$~Sct type (DSCT) are radial and non-radial pulsators which are found in the classical instability strip. They are of spectral type A2 to F5 and most of them are located on the Main Sequence (MS) of the Hertzsprung-Russell diagram or above it.  Pre-Main Sequence DSCT have also been discovered in young clusters or in the field \citep[][and references therein]{zwintz08}. 
DSCT  can pulsate in radial and non-radial pressure (p) modes,  and in mixed  pressure/low-order-gravity (p/g) modes, with a rich frequency spectrum   typically in the frequency range 4-50 \cd. The  main mechanisms driving the pulsations is the same as in the other variables in the classical instability strip,  the well known $\kappa$ mechanisms \citep{bk62}.  The mixed modes are caused by the change of the Brunt-V\"{a}is\"{a}l\"{a}  frequency in the stellar core as stars evolve, so that the same  wave can propagate as  a g-mode in the interior and as p-mode in the envelope \citep{aiz77}.  The interest of these modes is, therefore, that they allow to probe to a deeper extent the stellar structure.

The DSCT instability strip is partially overlapping  with that of $\gamma$~Dor  variables. These are  a group of MS  stars  a few hundred degrees cooler \citep{grigah10},  and pulsating in high order g-modes driven by convective blocking \citep{guz00} and with typical frequencies of  $\sim$ 1 \cd. In  recent years  pulsators showing both p-modes and  high order g-modes have been discovered (hybrid $\delta$~Sct / $\gamma$~Dor stars). The first is in the eccentric binary HD~209295 \citep{handler02} (even if a tidal origin of the $\gamma$ Dor pulsation was suggested afterwards), many other discoveries followed, especially thanks to space missions as CoRoT and {\it Kepler} \citep[][and references therein]{grigah10}. The further advantage of hybrid pulsators is that  for high order g-modes  the asymptotic approximation can be used \citep{sm07}.

 Precise stellar and atmospheric parameters are needed for the asteroseismic modeling of pulsating stars. SB2 
eclipsing binaries are best suited for such an analysis by combining the photometric with the spectroscopic data. Examples of 
such combined analyses of stars showing $\delta$\,Sct-like oscillations can be found e.g. in \citet{2011MNRAS.414.2413S}
in combination with Lehmann et al. (2013) or in \citet{2013arXiv1306.1819H}. \citet{2011MNRAS.414.2413S} find for
the short-period eclipsing binary KIC\,10661783 at least 68 frequencies, of which 55 or more can be attributed to pulsation modes.
Based on the analysis of the Kepler light curve, a mass ratio was derived that makes the system semi-detached and a suspected
oEA star (active Algol-type system showing mass-transfer and $\delta$\,Sct-like oscillations of the gainer, 
\citet{2002ASPC..259...96M,2003ASPC..292..113M}). Combining the photometric findings with the spectroscopic analysis of the SB2 star
revealed, however, that the star must have a detached configuration which makes it a post-Algol type system of still extremely small mass
ratio (Lehmann et al. 2013).

\citet{2013arXiv1306.1819H} combined Kepler photometry and ground based spectroscopy of KIC 4544587, a short-period eccentric 
eclipsing binary system showing rapid apsidal motion. Their combined analysis delivers the absolute masses and radii of the 
components and shows that the primary and secondary components of KIC 4544587 reside within the $\delta$\,Sct and $\gamma$\,Dor 
instability regions, respectively. The important result is that the 31 oscillation frequencies found in total can be attributed, besides to 
self-excited p and g modes, also to tidally excited g modes and tidally influenced p modes.

In this paper we present the analysis of the bright \kepler target  KIC~3858884 ($K_p=9.227$, $V=9.28$), an eclipsing binary whose   light curve is   characterized by deep eclipses and  an additional periodic  pattern  suggesting the presence of a $\delta$ Sct pulsator.  
The system  has an orbital period of  about 26 days, but  binarity  affects  also the out-of-eclipse light curve because of the high eccentricity  ($e \simeq 0.5$). 
What makes  this system different from the  above-mentioned cases  is that eclipses and pulsations are of similar strength and strictly interlaced, since during  eclipse only a small fraction of the pulsating star is visible.  This required the development of a new disentangling procedure to take into account, at least at first order,  the effect of eclipses on the observed pulsation pattern.
 
The target was object of a follow-up  campaign to gather high resolution spectra covering the full orbit. A preliminary analysis, based on the first \kepler data releases and 18 spectra from the Bohyunsan Observatory Echelle Spectrograph (BOES) was presented by \citet[hereafter L12]{lee12}.
The system turned out to be a double-lined spectroscopic binary (SB2), but the limited orbital coverage of the spectra and the relatively short length  \kepler photometry available at the time  did not allow an accurate determination of  system parameters. 

This paper is organized as follows: Sects.~\ref{phot} and \ref{spectro} describe the available data and their reduction; Sect.~\ref{span} is devoted to the analysis of the disentangled component spectra, providing the atmospheric parameters;   Sect.~\ref{lc-rv-an}  deals with the light and radial velocity curve analysis;    Sect.~\ref{puls}  analyzes  the pulsational properties of the  components and, finally, Sect.~\ref{abspar}  presents a  comparison between the physical elements derived from the analysis and stellar evolutionary/pulsation  models.

   \begin{figure}[t!] 
  \centering
   \includegraphics[width=9.cm, trim= 20 10 0 0]{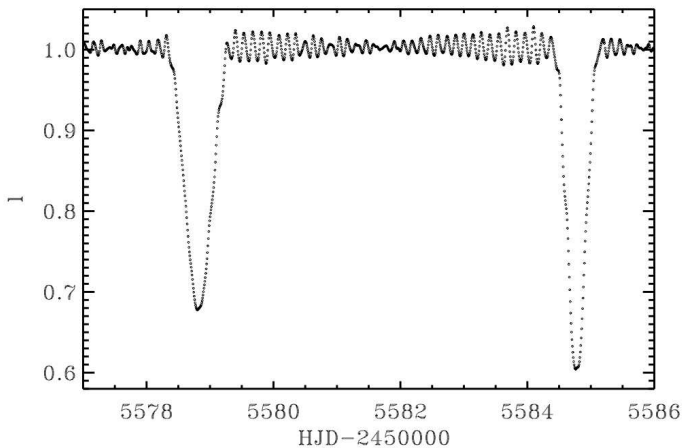}
      \caption{ Blow-up of a section of the light-curve  in Fig.~\ref{lc}.}
        \label{lc_blow}
          \end{figure}

\section{\kepbf photometry}
\label{phot}

\eb has been in the  \kepler target list since the beginning of the mission. It has been observed both in the ``short cadence'' (SC) and ``long cadence'' (LC) mode; in these modes the original integration times of aboout 6$^\mathrm{s}$  are summed on board and yield a final integration time of,  respectively, 59$^\mathrm{s}$ and 29.4$^\mathrm{m}$ \citep{gilliland10}. Short cadence is available for a small number of targets (0.3 \% of the  total), therefore for each run targets are inserted in or dropped from the SC observation list according to priority criteria. Also the LC mode has its priority list, as a consequence a given target can be observed only at some epochs.  Kepler data are delivered in ``Quarters'' (Q0, Q1, ..., Qn), typically three months long (a quarter ended when the spacecraft rolled to re-align its solar panels). \eb was observed in the SC mode during
  quarters Q2, Q8 and Q9 (a total of 358307 observed points) and in the LC  one during quarter  Q0, Q1, Q2, Q3, Q4, Q8, Q11, Q12, Q13 and Q15 (39949 points). Because the  eclipses of \eb last  between 16 and 18 hours and the dominant pulsation frequencies correspond to periods of a few hours,  we used  for our analysis mainly the short cadence data which assure a dense sampling at crucial orbital phases (and a higher Nyquist frequency).
  
   The Simple Aperture Photometry (SAP) data from the MAST Archive\footnote{http://archive.stsci.edu/kepler/} 
   were cleaned of obvious outliers and normalized to the median value of each quarter, by using the Pyke  tools \citep{still12}. Q2 data cover 89 days, Q8 and Q9 data 170 days,  for a total of about 10 orbital cycles.
   A section of this  light curve corresponding to Q8 and Q9 quarters   is shown in Fig.~\ref{lc}, the data were re-binned  to 600$^{\mathrm{s}}$ for better visibility. The blow-up of Fig.~\ref{lc_blow} shows the pulsations   and eclipses in greater detail.
    
Thanks to the results of L12  we had already  preliminary values for the binary period and eccentricity  (P$_{\mathrm{orb}}=25\fd9521$ and $e=0.46$).  Their binary model, however, is  from fitting a binary-only light curve which was  obtained by  pre-whitening the original one with {\it all} the  frequencies detected in its out-of-eclipse sections.  This is certainly an over-correction, because the light curve of a  system with high eccentricity, fractional radii, and orbit orientation as those obtained in L12 will show  binary signatures  out of eclipse as well. Stronger surface distortion and  enhanced proximity effects at periastron are at the origin of the typical bump appearing in the light curve of  eccentric systems in the shorter phase interval between eclipses.  As a consequence, the harmonic analysis, though restricted to the out-of eclipse sections, will  still contain the orbital frequency and its overtones. These frequencies belong to the binary 
signal and should not be pre-whitened as  they are not related to pulsation.

To estimate the relevance of this effect  we computed with PHOEBE \citep{prsazw05} a binary model with the parameters of L12. The resulting synthetic light curve (after adding gaussian noise with  standard deviation from their pre-whitened binary light curve) was analyzed in the out of eclipse sections with PERIOD04 \citep{p04}. The evidence was that the eccentricity bump, in that case, yields significant orbital overtones up to $ 15 f_{\mathrm{orb}}$.  
Therefore, in the following, after performing the harmonic analysis with both PERIOD04 and SIGSPEC \citep{sigspec}, we excluded from pre-whitening the orbital frequency overtones up to the value found in the simulation.
  
     \begin{figure}[ht!] 
  \centering
   \includegraphics[width=8.5cm,trim=15 0 0 0]{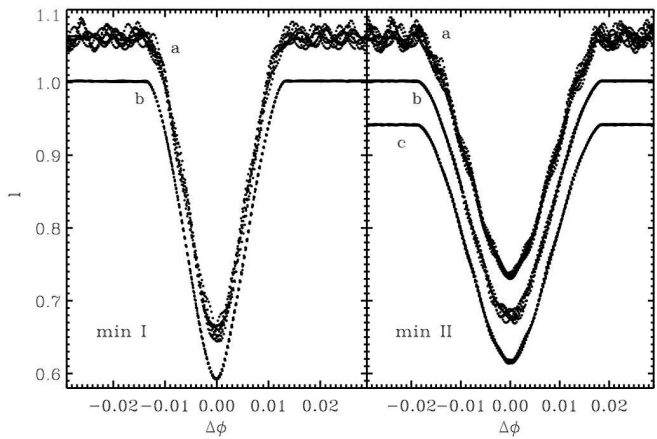}
      \caption{ Sections of the phased light curve around minima. Left panel: primary minimum, right panel: secondary  minimum. $\Delta \phi$ is the phase difference with respect to conjunction phase. The curve labeled with ``a"  are derived by phasing   LC0;  curves ``b''  after  pre-whitening with pulsations, with no correction for visibility during eclipse. Curve ``c'' of the right panel is pre-whitened taking visibility into account (see text).}
        \label{minblow}
          \end{figure}

         The disentangling of the EB signal from pulsations is typically performed with an iterative procedure, see e.g. \citet[][hereafter MMG13]{cm13}. The process consists in: 
 1) pre-whitening the original light curve (LC0)  with the frequencies from the harmonic fit computed outside eclipses (if necessary after replacing the eclipse points with a local fit),   
 2) subtracting  from LC0 the binary model obtained by fitting the residuals of the previous step, 3) pre-whitening again LC0 with the harmonic fit, determined this time on the full curve;
  the last two steps are iterated until convergence.

         This procedure was, for instance,  successfully applied in the analysis of  CoRoT~102918586 (MMG13), a grazing eclipsing binary with a $\gamma$ Dor component whose pulsation amplitude  is   comparable to the (single) eclipse depth. 
  In the case of KIC~3858884, however, a large fraction of the eclipsed star is out of view during eclipse,  being the orbital inclination  close to 90 degrees ($88.8\degr$ in L12). As a consequence the pulsations computed out of eclipse cannot be subtracted ``as are'' at all phases. This is evident from Fig.~\ref{minblow} showing a blow-up of both minima of the phased light curve before (curves labeled with  ``a'')  and after (curves ``b") subtraction of the pulsations,  whose computation is described later in this section. The ephemeris used for phase computation of the curve is given in Eq.~\ref{ephe}.
  
  From  inspection of the top curve  of each panel  of Fig.~\ref{minblow},   we conclude that the pulsating star is the secondary component or, more precisely, that the greatest periodic luminosity variations belong to the secondary star, since the dispersion due to pulsations of the folded curve significantly decreases only during  secondary minimum (Min~II).
 The curves ``b''   show the effect we mentioned before:  if the visibility of the pulsating star is not taken into account the mere subtraction of the harmonic fit from out-of-eclipse sections increases the scatter  around secondary minimum with respect to that of the original curve.  The opposite is true for the primary minimum, as in that case the pulsating star is in full view.

  In light of these evidences we modified the disentangling procedure as follows: the first harmonic analysis was performed  on LC0 after replacing the eclipses with a local fit of the neighboring points and gaussian noise.  This yielded about 300 frequencies (with a significance criterion of the ratio of amplitude to local noise  $S/N >4$, according to the criterion proposed by \citet{breger93}).
  We then subtracted  from LC0 the harmonic fit, excluding the orbital overtones up to 18 $ f_{\mathrm{orb}}$, and computed with PHOEBE a preliminary binary model, just to estimate for each point  the fraction of light from the pulsating star at that time with respect to the median value. We assumed that also the amplitude of pulsations during eclipse is reduced by this factor, and weighted accordingly the harmonic fit to be subtracted from LC0.
  This is only a rough estimate, because the actual value of this fraction  certainly  depends on the pulsation mode and the spherical wave-numbers, $\ell$ and $m$,  (sectorial modes will act differently from tesseral or axisymmetric modes), but even this first order treatment is sufficient for a meaningful decrease of the dispersion (see the lower curve in the right panel of Fig.~\ref{minblow}, labeled with ``c'').

         \begin{figure}[ht!] 
  \centering
   \includegraphics[width=8.5cm, viewport = 40 10 495 350, clip=true] {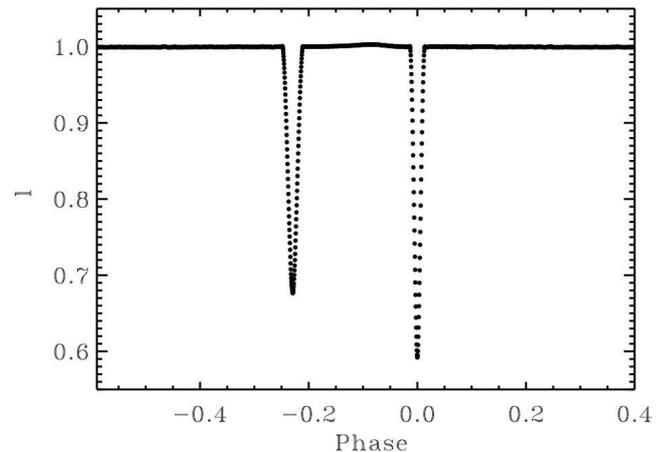}
      \caption{ The phased light curve of the eclipsing binary as  residuals of the corrected harmonic fit, orbital overtones excluded.  Phases are derived using the ephemeris of Eq.~\ref{ephe}.}
        \label{lcph}
          \end{figure}

The phased final curve due to only to the orbit is shown in  Figure~\ref{lcph}. For the sake of brevity we show only the final outcome of the disentangling procedure (after two stages of pre-whitening of LC0). To reduce the computing time in binary modeling,  the input light curve was phased and  further binned with a variable step (namely we computed normal points  in phase bins of 0.0004 and 0.001 in and out of eclipse). The total number of points of the curve is then reduced to 1106.  On the other hand the harmonic analysis and pre-whitening was performed  at first on the full time series,  when it became evident  that there was no power in the Fourier decomposition at frequencies higher than 50 \cd, we binned also this time series to a time interval of  600$^\mathrm{s}$.
  
An accurate  orbital period  was derived by analyzing the pre-whitened time series with  the Analysis of Variance (AoV) algorithm of \citet{AoV89}, as implemented in the VarTools program \citep{hartman08} following \citet{devor05}. For this computation we used also the LC curves, which cover a longer timespan  (1153$^{\mathrm{d}}$ or 44 orbital cycles).  

The resulting  ephemeris is:
\begin{equation}
  T_{\mathrm{minI}} = 2455013.8311 (1) + 25\fd95200 (5)  \times E.
\label{ephe}
\end{equation}
\section{The spectroscopic follow-up of \eb\label{spectro}}

A follow-up campaign with different instruments was necessary  to get high-resolution spectra covering 
the long orbital cycle  with a good sampling. The observations were performed with the
Coud\'e echelle spectrograph at the 2-m telescope of the Th\"uringer Landessternwarte (TLS) 
Tautenburg, (Germany), with the Bohyunsan Optical Echelle Spectrograph (BOES) at the 1.8m 
telescope at Youngcheon (South Korea), with the HERMES spectrograph at the 1.2m  Mercator 
telescope at the Roque de los Muchachos Observatory (La Palma, Spain), and with the HRS spectrograph
at the 9.2m Hobby-Eberly telescope (HET) at the McDonald Observatory (US).  The journal of observations
is given in Table\,A.1 that also lists the measured radial velocities (RVs).

The spectral ranges and resolving power are 472-736\,nm, R=32\,000 for the TLS, 380-900\,nm, 
R=85\,000 for the HERMES, 350-1050\,nm, R=30\,000 for the BOES,  and 440-585 plus 610-780\,nm, R=30\,000 for the HET spectra.
All spectra were reduced using the following steps:
cosmic ray hits removal, electronic bias subtraction, flat-fielding, electronic bias and straylight subtraction, spectrum extraction,
2-D wavelength calibration, and normalization to the local continuum.
Standard ESO-MIDAS packages and a special routine to calibrate the 
instrumental zero point in radial velocity by using a large number of telluric O$_2$ absorption lines
were used for the TLS spectra.
The HERMES and HET spectra were reduced using the corresponding spectrum reduction pipelines \citep{raskin11, mack13} and the BOES spectra
using the  IRAF tools.

The average signal to noise ratio of the spectra from the different instrument is: S/N=160   (TLS),  S/N=120  (HET),  S/N=170   (BOES), and  S/N=44   (HERMES).
 
\section{Spectroscopic analysis\label{span}}
\subsection{Spectrum decomposition \label{spdec}}

\begin{table}[t!]\centering
\caption{Abundances of the binary components.  The second column gives the atomic number, the third the assumed solar abundance, the fourth and fifth the 
difference with respect to the solar value (see text).}
\label{chemcomp}
\begin{tabular}{lrrll}
\hline\hline
 Element& Z & Sun	& Primary & Secondary \\
\hline \\ [-0.3cm]
 	C	&  6 & $-$3.65   &                    ~~~~~$-$			& $- 0.34 ^{+0.37}_{-0.78} $ \\[0.1cm]
	Mg	&12 & $-$4.51   &  $-0.37 \pm0.47 $                   & $-0.37^{+0.34}_{-0.46}  $ \\   [0.1cm]
	Ca	&20 & $-$5.73   & $-0.44\pm0.30$    		& $ -0.20 \pm 0.24$ \\   [0.1cm]
	Sc	&21 & $-$8.99   &   $-1.11 \pm 0.66 $   		& $+0.12\pm0.33 $ \\ [0.1cm]
	Ti	&22 & $-$7.14   & $-0.37 \pm 0.20 $   		& $-0.27 \pm 0.17 $ \\ [0.1cm]
	Cr	&24 & $-$6.40   &$-0.04 \pm 0.16   $  		& $-0.24 \pm 0.17 $ \\     [0.1cm]
	Mn	&25 &  $-$6.65  &$-0.16 \pm 0.36  $  		& $-0.27 \pm 0.35 $  \\ [0.1cm]
	Fe	&26 & $-$4.59   &$-0.09 \pm 0.07 $   		& $-0.26 \pm 0.07 $ \\ [0.1cm]
	Ni	&28 & $-$5.81   &$+0.17 \pm 0.13$			& $-0.13 \pm 0.14 $	 \\ [0.1cm]
	Y	&39 & $-$9.83	  &$+0.72 \pm 0.25 $			& $+0.53 \pm 0.24 $   \\ [0.1cm]
	Zr	&40 & $-$9.45   & $ +0.80^{+0.39}_{-1.16} $ 	& $ +0.71^{+0.41}_{-1.26}$ \\ [0.1cm]
	Ba	&56 & $-$9.80   &$ +1.14^{+0.55}_{-1.02}  $  	& $ +0.97^{+0.59}_{-0.97}  $  \\[0.1cm]
 	La	&57 & $-$10.91 &$+0.94^{+0.28}_{-0.39}   $  	& $+0.54^{+0.38}_{-1.00}   $ \\[0.1cm]
	Ce	&58 & $-$10.46 &$+0.74^{+0.33}_{-1.03}$ 	& ~~~~$ -                     $\\[0.1cm]
	Nd	&60 & $-$10.39 &$+0.56^{+0.23}_{-0.36}$ 	& $+0.21^{+0.35}_{-1.32}$\\[0.1cm]
\hline
\end{tabular}
\end{table}  

For the analysis of the spectra of the components, we decomposed the observed, composite spectra using the Fourier
transform-based KOREL program \citep{korel3,korel4}. This program optimizes the orbital elements together with
the shifts applied to the single spectra of the components to build the mean decomposed spectra. 

 Whereas we used all the spectra for the measurement of the RVs, only the TLS spectra were included 
to derive the decomposed spectra. The reason is that the HERMES spectra are of much lower S/N (due to small exposure times), the HET spectra show
badly normalized Balmer line profiles, and the BOES spectra suffer
from a slight ripple in the continua which causes problems in the continua of the decomposed spectra as described below.

There are two drawbacks
of the program. One comes from the usage of the Fourier transformation. The zero frequency mode in the Fourier domain 
is unstable and cannot be determined from the Doppler shifts alone \citep{hensberge2008}. This effect gives rise to
low-frequency undulations in the continua of the decomposed spectra and prevents in many cases 
from an accurate determination of the local continua.  The undulations are enhanced when broad lines like Balmer lines
occur in the spectral region of interest or when the input spectra already show slight undulations in their continua.
The problem can be overcome when the line strengths of the stars vary with time and the degeneracy of the low frequencies
in the Fourier domain is lifted.

The second problem is common to all methods of spectral
disentangling, namely that the produced decomposed spectra are normalized to the common continuum of the composite input
spectra. For the spectrum analysis, however, we need the spectra normalized to the continuum of the single components.
Such a renormalisation can only be performed when the flux ratio between the components is known.  In a first attempt,
we computed with PHOEBE the expected luminosity ratio in the Str\"{o}mgren $b$ and $y$ bands  for the stellar parameters derived
in the light curve solution.
These two passbands cover or enclose the wavelength range of our spectra.  However, the decomposed spectrum of the 
secondary component showed after the normalization line depths larger than unity (or normalized fluxes below zero) so that
we must assume that the flux ratio derived from the light curve was too small. We, therefore, included the determination
of the wavelength dependent flux ratio into the spectrum analysis by using it as an additional free parameter, see 
\citet{lehm13} for a description of the method.
 
 \begin{figure}[t!]
   \includegraphics[angle=-90, width=8.9cm,  trim= 10 40 170 0]{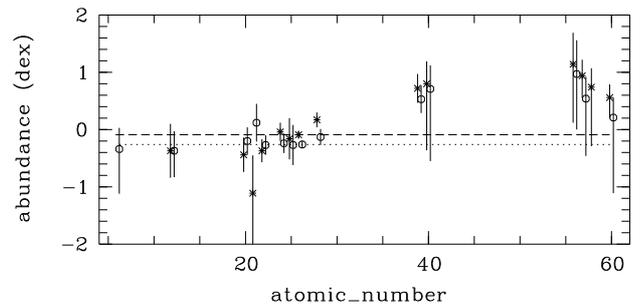}
      \caption{Abundances of the primary (asterisks) and secondary (open circles) components in dex relative to the solar values.
      The dashed and dotted lines mark the [Fe/H] of the primary and secondary components, respectively.}
      \label{abundances}
          \end{figure} 

{ For the spectral disentangling with KOREL, we used the wavelength range 472 to 576\,nm, from the blue edge of the TLS spectra
to the wavelength where stronger telluric lines become visible. In the solution, we fixed the
orbital period to the value obtained from the LC analysis which is much more precise than we can derive from the spectra
alone. In the program setup, we allowed for variable line strengths. In a first step, we computed two solutions, the first 
one by including the spectra taken during primary minimum and a second one without these spectra.
The comparison showed that in the first solution the decomposed spectrum of the primary was slightly broadened by the strong 
Rossiter-McLaughin effect (hereafter RME). The second solution, on the other hand, showed a strong undulation of the continuum of both decomposed spectra.
The reason can be found in the above mentioned lifting of the degeneracy of the low Fourier frequencies by the variable line strengths
which does not work anymore when excluding the primary eclipse from the computations. Finally, we found a compromise by excluding
all but two spectra taken at Min\,I. The resulting decomposed spectra did not show any undulations anymore and no difference in
the line profiles compared to the exclusion of all spectra around Min\,I could be found.

\subsection{Spectrum analysis}
\label{spectr_anal}
 The decomposed spectra of the components were analyzed using the spectrum synthesis
method. This method compares the observed with synthetic spectra computed on a grid in atmospheric parameters and uses
the $\chi^2$ of the O-C residuals as the measure of the goodness of fit see  \citet{2011A&A...526A.124L} for details.
We used the SynthV program \citep{1996ASPC..108..198T} to compute the synthetic spectra based on a library of
atmosphere models calculated with the LLmodels code \citep{2004A&A...428..993S}. Both programs are non-LTE based
but provide the opportunity to vary the abundances of single chemical elements.

 As mentioned in before, we added the flux ratio as a free parameter to the spectra analysis. It means that
the analysis had to be performed on both decomposed spectra simultaneously, searching for the best fit in the
atmospheric parameters of both stars and the flux ratio between them.}
Free parameters of our fit are effective temperatures \tef, metal abundances [M/H], microturbulent velocities 
$v_{\rm turb}$, and projected equatorial velocities \vsin. Because of its ambiguity with \tef\ and [M/H],
we fixed the \lgg\  of both components to the values obtained from the LC analysis. We started with
scaled solar abundances for the [M/H] of both stars. After finding the best solution in all parameters, we 
re-normalized the decomposed spectra according to the derived flux ratio. Then we fixed all the parameters to the 
derived ones and repeated the analysis for both spectra separately by varying the abundances of all chemical
elements where contributions could be found in the spectra, starting with the abundance tables corresponding to the
derived [M/H]. In a next step, after optimizing all individual abundances, we repeated the analysis of both spectra 
as described above based on the modified abundance tables. Changes in the atmospheric parameter values have been  
marginal so that we stopped at this point.

Table\,\ref{chemcomp} lists the results of abundance analysis, Table\,\ref{atmos} gives the final atmospheric parameter
values. The second column of Table\,\ref{chemcomp} gives the atomic number, the third the assumed solar abundance as 
$\log(N_{\rm el}/N_{\rm total})$, corresponding to  \citet{2005ASPC..336...25A}. The fourth and fifth columns list
 the difference with respect to the solar value (+ means enhanced).
The parameter errors represent 1$\sigma$ errors. The atmospheric parameters were derived from
the full grid of parameters but for both components separately. The ambiguity between the parameters are included
in the error analysis in this way per star.
 We did not have enough computer power to include the combined effects from both stars into the error calculation
  as they may arise during the combined analysis when determining the flux ratio, e.g. to consider the effect of 
 changing $T_{\rm{eff,1}}$ on the parameter values of the secondary component.
 For the abundances, we included the \tef of the corresponding star into the error analysis of the single elements.

Figure\,\ref{abundances} shows the abundances sorted by their atomic number. The derived abundances of both components 
agree within their 1$\sigma$ errors for all but three elements. The outliers are Fe and Ni, the elements with the smallest
abundance errors, and Sc.
In the case of Fe and Ni, this could be a hint that the errors were underestimated. In the case of Sc, on the other hand, 
the difference of 1.23 dex can be confirmed from a visual inspection of the decomposed as well as of the observed, 
composite spectra.

Because of the ambiguity between \tef and abundances it is important to note that the abundances given in Table \ref{chemcomp} 
are based  on the spectroscopically derived \tef. They are part of a consistent spectroscopic solution, but  may vary if a different 
\tef (such as the value derived from the combined light and radial velocity curve solution, see Sect.~\ref{lc-rv-an}) is assumed.

The distribution of the abundances with atomic numbers of the primary component resembles that of BD\,+18\,4914 which was 
discussed in \citet{2011ApJ...743..153H} to be an Am star. Am stars are chemically peculiar early F to A stars. They are 
characterized by slow rotation (cut-off of about 100\,\kms), under-abundance 
of C, N, O, Ca and Sc, and overabundance of the iron peak and rare earth elements (see e.g. \citet{2008CoSka..38..129I}
for an overview). Our two stars show not too fast rotation and 
an overabundance of the heavy elements but not of those of the iron peak. Only one of the stars, the non-pulsating
primary component, shows a significant under-abundance of Sc.

\begin{table}[t!]\centering
\caption{Atmospheric parameters of the binary components.}
\label{atmos}
\centering
\tabcolsep 1.8mm
\begin{tabular}{lll}
\hline\hline
Parameter &   Primary & Secondary \\
\hline \\ [-0.3cm]
\tef  (K)                        &    6810$\pm$70 & 6890 $\pm$80 \\ 
\lgg  (cgs)                   &    3.6 fixed          & 3.7 fixed \\ 
\vsin           (\kms)          & 32.2$\pm$1.5& 25.7 $\pm$ 1.5\\
$v_{\rm turb}$  (\kms) & 3.72 $\pm$ 0.16 &  3.74 $\pm$ 0.60\\
\hline
\end{tabular}
\end{table}

\section{Light  and radial velocity curve analysis
\label{lc-rv-an}}   

The light and radial velocity curve solutions were performed with the current (``stable") version of PHOEBE.

We preferred a non-simultaneous solution of light and radial velocity curves, since the RV data are acquired at 
different epochs with respect to photometry and  the two data  sets are very different in terms of observed point 
number and accuracy. On the other hand the two solutions were connected by keeping fixed in each of them the 
parameters better determined by the other type of data.

\subsection{Radial velocities and preliminary orbital solution}
\label{rvs}
 We used two different programs, KOREL and TODCOR \citep{1994Ap&SS.212..349M,1994ApJ...420..806Z}, 
to measure the radial velocities (RVs), and  PHOEBE  to 
derive the orbital parameters.  This allowed to  cross check  the results but also to profit from the advantages of each method:
 KOREL provides disentangled spectra and  performs better in the analysis of spectra  at conjunction phases,
 while TODCOR is simpler and yields a more straightforward and consistent determination of uncertainties.
In both cases the included spectral  range  was 472-567\,nm,  which does not contain 
broad Balmer lines and is almost free of telluric lines.
 \begin{table}[t!]
\caption{Orbital parameters derived with KOREL (K) and from the cleaned KOREL (K+P) and 
TODCOR (T+P) RVs using PHOEBE. Errors are given in units of the last digits in
parentheses.}  
\label{orbelements} 
\centering                   
\begin{tabular}{lrrr}        
\hline\hline
& \multicolumn{1}{c}{K} & \multicolumn{1}{c}{K+P} & \multicolumn{1}{c}{T+P}\\                   
\hline
$a$ ($R_\odot$)    & 56.94  & 57.08(16)  & 57.22 (20)\\
$e$                & 0.4661 & 0.4674(15) & 0.4686(24)\\
$\omega$ (deg)     & 21.87  & 21.90(23)  & 21.48(35)\\
$q$                & 0.9996 & 0.9991(54) & 0.9880(68)\\
$T$ (2\,455\,000+) & 11.974 & 11.9804(86) & 11.9795(14)\\
$\gamma$ (\kms)    & &           &16.14(13)\\
\hline
\end{tabular}\\
\end{table}
\begin{table}[t!]
\caption{Frequencies found in the residuals of the RV fit of the secondary component, 
based on the RVs derived with KOREL and with TODCOR. The frequencies found in the Kepler light
curve are given for comparison.}   
\label{freqtab}
 \centering  
\begin{tabular}{lclr}       
\hline\hline
&  &$f$ (d$^{-1}$)&	\multicolumn{1}{c}{amplitude}\\ 
\hline
KOREL  &$f_1$ & 7.23074(6) & 2.6(2) kms$^{-1}$\\      
       &$f_2$ & 7.47340(6) & 2.9(2) kms$^{-1}$\\      
TODCOR &$f_1$ & 7.23056(6) & 3.4(2) kms$^{-1}$\\      
       &$f_2$ & 7.47341(6) & 3.4(2) kms$^{-1}$\\      
Kepler &$f_1$ & 7.2306(1)  & 10.2(2)$\,\cdot10^{-3}$\\
       &$f_2$ & 7.4734(1)  & 9.1(2)$\,\cdot10^{-3}$\\
\hline
\end{tabular}
\end{table}
KOREL  delivers the velocity shifts  with respect to the composite spectrum and, directly, the searched 
orbital parameters. As first step we obtained a preliminary orbital solution, based on all the out-of-eclipse spectra, which is given in 
Table\,\ref{orbelements}, column K. In the following, we used  all spectra, including those at eclipse phases.  Since KOREL 
does not provide straightforward error estimates on the derived parameters (nor on the RV shifts) and does not consider the RME and proximity 
effects, a  fit of the  KOREL RV curves was performed with PHOEBE, in order to independently obtain the orbital parameters  together with their errors.

\begin{figure}[t!]
   \centering
   \includegraphics[width=8.5cm,bb=20 50 504 608]{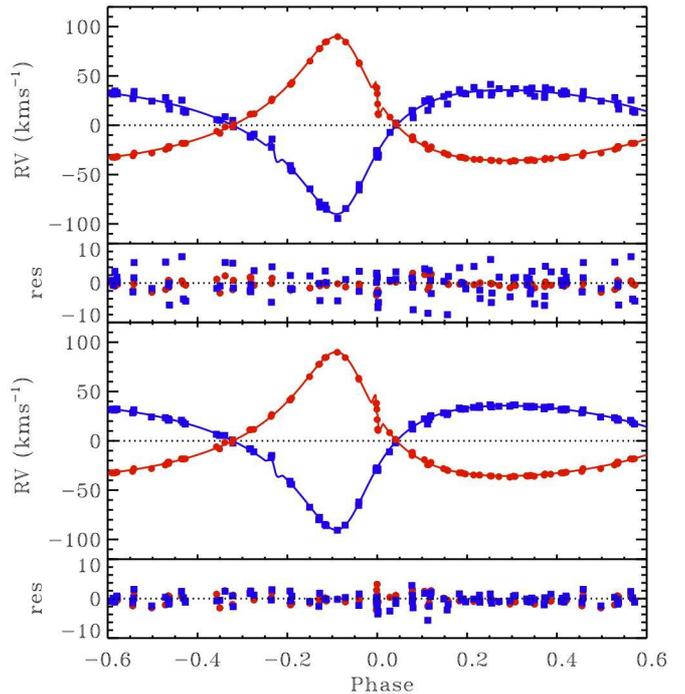}
         \caption{Upper two panels: the phased RV shifts of  \eb components from KOREL. 
	 Filled circles: primary star. Filled squares: secondary star. Curves: PHOEBE fit.  
	 Second panel: fit residuals, with the same symbols.  Lower two  panels: the same 
	 after prewhitening with the two significant frequencies found in the residuals.}
         \label{rvc}
   \end{figure}

PHOEBE calculates the orbit simultaneously fitting  the RV curves of both components and
includes the RME,  though only in the hypothesis of alignement of  rotation and orbit axes, and the proximity effects (on the basis of a model of the 
surface brightness distribution from the light curve fit). 
The adjusted parameters were  the system semi-axis, $a$, the eccentricity, $e$, the longitude of periastron, $\omega$, the mass ratio, $q$, 
 the  barycentric velocity, $\gamma$, and the origin of time in PHOEBE, from which we derived the epoch of periastron passage,
 $T$.  Figure \ref{rvc} shows the KOREL shifts and the  PHOEBE best fit.

We subtracted the resulting fit from the observed RVs
and performed a period search in the  residuals with PERIOD04. For the secondary component, we found the 
same two frequencies that  dominate 
the oscillations in the Kepler light curve (see  Tables~\ref{freqtab} and \ref{freqs}). No periodic short-term variations were instead found for the primary. 
Finally, we obtained  the final fit with PHOEBE, after pre-whitening the secondary RV curve with the detected frequencies.
 The results are listed in Table\,\ref{orbelements} in column K+P.
The listed uncertainties  are the formal errors of the  fit  and are certainly underestimated,  because of their same origin 
(see Sect.~\ref{uniq}) and of the lack of  errors of the  KOREL RVs.

\begin{figure}
\includegraphics[angle=-90, width=8.8cm]{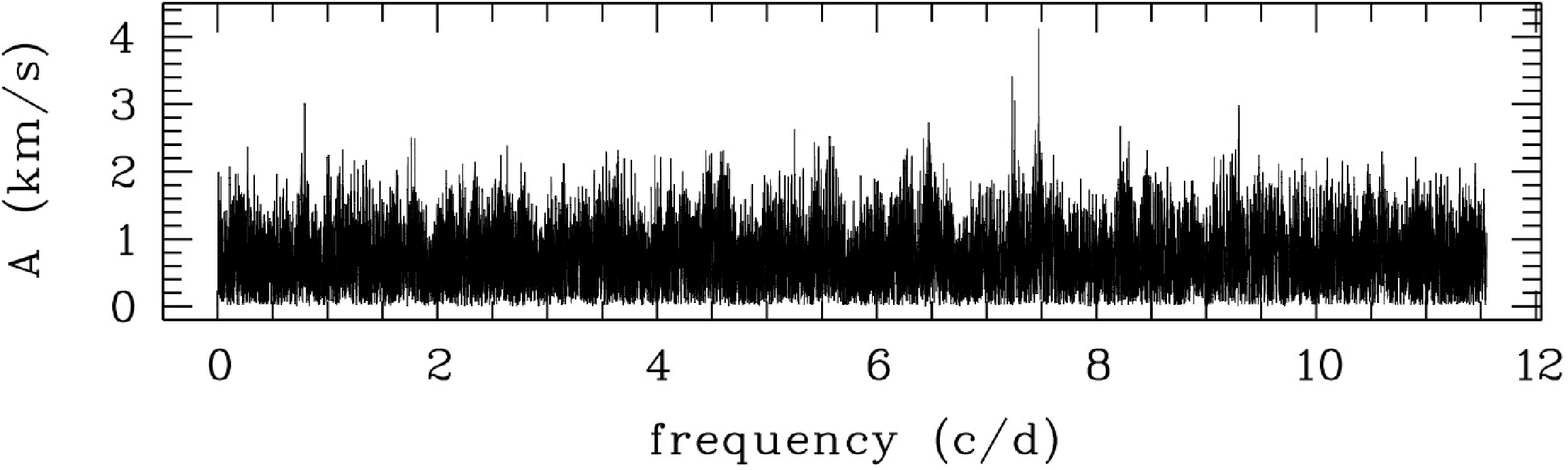},
\includegraphics[angle=-90, width=8.8cm, trim=60 0 0 0 0]{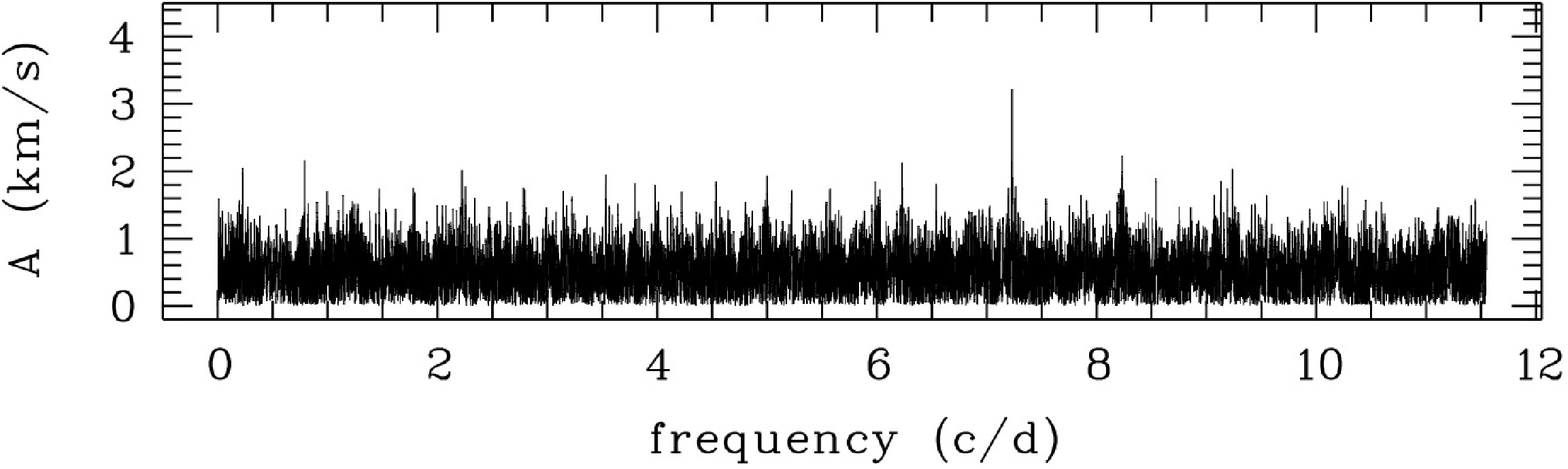}
\includegraphics[angle=-90, width=8.8cm, trim=60 0 0 0 0]{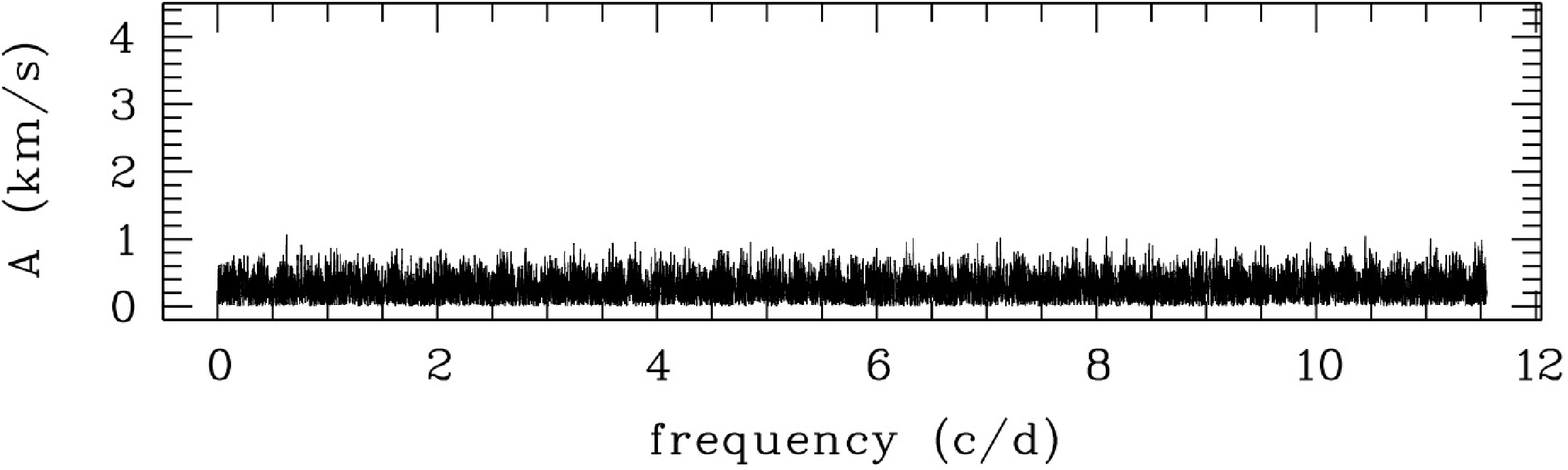}
\caption{ Amplitude spectra, based on our complete dataset of 83 spectra. From top to bottom: Original data, largest peak $f_1$;
$f_1$ subtracted, largest peak $f_2$; $f_1$ and $f_2$ subtracted.}
\label{fft}
\end{figure}

 Because the described procedure is a mixture of two different methods - the RVs are determined in a procedure 
 not accounting for the RME and proximity effects whereas the calculation of the orbit with PHOEBE includes them -
 and because of the  problems to get a sound  estimate of the errors,  we redetermined the RVs from the observed spectra
 by using  the implementation of the TODCOR  algorithm by one of the authors (HL).

TODCOR is applied to the observed composite spectra and performs a two dimensional cross-correlation for the
contributions from each  star, using two different templates. We used as templates  two synthetic spectra based on 
the parameters obtained for the two stars from spectrum analysis
(see Sect.\,\ref{span}). The resulting RVs of both components are listed in Table\,\ref{rv} of the 
Appendix.
Also in this case we  computed  the spectroscopic orbit with PHOEBE, performed the analysis of the residuals with PERIOD04
finding the same two frequencies as from the KOREL RVs (Table~\ref{freqtab}), and derived the final fit after pre-whitening them.
Fig.\,\ref{fft} shows the finding periodograms. The final orbital parameters are listed in Table\,\ref{orbelements}, column T+P.

The comparison of the results obtained from the KOREL and the TODCOR RVs shows that they are of about the same accuracy.
All orbital parameters agree within 1$\sigma$.
Besides,  all the  derived frequencies of the short-term variations are in perfect agreement and with the photometric ones. Taking both 
parameter sets together, we can say that the most important parameter, the mass ratio, is of $q$=0.993$\pm$0.012.
 This is a conservative estimation that assumes that the two methods are of comparable accuracy. 

         \begin{figure}[hb!] 
  \centering
   \includegraphics[width=9cm,  trim= 50 10 10 10]{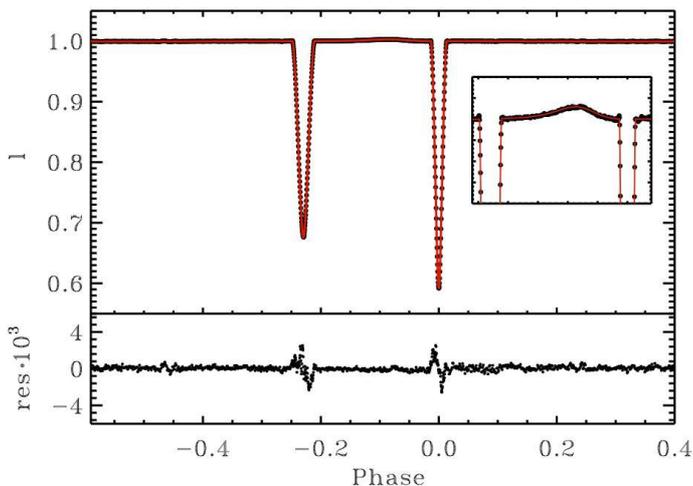}
      \caption{ Upper panel: the phased light curve of the eclipsing binary and the  final fit with PHOEBE (full line, in red in the on-line version),  inset: a blow-up around periastron phase showing the fit  around  periastron phases.  Lower panel: fit residuals. }
        \label{binlc}
          \end{figure} 
 
 In Fig.~\ref{rvc} we preferred to show the RV shifts from KOREL, because that method provides a few more points around primary minimum, 
 allowing to better trace the Rossiter-McLaughin effect.
  For the combined photometric/spectroscopic solution, however, we will favour the TODCOR-based result as  input. The reason is that 
its derivation seems to be more consistent, having in mind that the KOREL solution is based on RVs derived without 
considering the Rossiter-McLaughin and proximity effects, computed with PHOEBE to get the errors of the parameters.

The inspection of the radial velocity curves confirms that the dominant pulsations belong to the secondary component,
since the RME, visible in both curves, indicates that the component showing pulsation-related scatter
in  velocity is the one eclipsed at phase $\sim 0.77$ (or -0.23). On the other hand neither from the light curve nor from spectra we can
deduce that all the components of the periodical pattern detected in the light curve belong to the same star.

The in-eclipse RV poits are too few for modeling RME and deduce information on the  rotation axis alignment,   the apparent simmetry around the unperturbed curve of the four points at primary minimum
are consistent with a parallel rotation axis of the primary component.
\subsection{Light curve solution}
\label{lcsol}

In the light curve solution  we  adjusted  the inclination $i$,  the secondary effective temperature, $T_{\mathrm{eff,2}}$, and the surface potentials $\Omega_{1,2}$; the primary passband luminosity $L_1$ was separately computed  rather than adjusted,  as this allows for a smoother convergence to the minimum. The  mass ratio $q$, was instead  fixed to the values derived  from the fit of the radial velocity curves (from TODCOR parameter set);  the eccentricity, $e$, and the longitude of periastron, $\omega$, were initially fixed to the spectroscopic values and only finely tuned in the light curve fit.
 \begin{table}[t!]
 \caption{Orbital and physical parameters of \eb}
 \label{sol}
 \tabcolsep 1.7mm
 \centering
 \begin{tabular}{lccc}
 \hline\hline
			&					&  System		&                      \\
                    			& Primary				&			&  Secondary         \\
\hline \\ [-0.3cm]
$i$ ($^\circ$)			&	 	& $ 88.176 \pm 0.002$   &                      \\
$e$			   		&		&$0.465  \pm 0.002$	&			\\
$\omega$				&		& $21.61^\circ  \pm 0.01$		&			\\
$q$					&		&$0.988	 \pm  0.02	$&			\\
$a (R_{\odot})$			&&		$	57.22  \pm  0.22$	&	\\

$T_\mathrm{eff}$ {(K)}		& $6800^{a } \pm  70$				&  		& $6606 \pm 70$		\\
$M  (M_{\odot})$                			&  $1.88 \pm 0.03$         &                & $1.86\pm 0.04$	\\
$R  (R_{\odot})$                			&  $3.45 \pm 0.01$         &                & $3.05\pm 0.01$	\\
$\logg$                		           	&  $3.63  \pm 0.01$          &                & $3.74 \pm 0.01$	\\
\hline
\end{tabular}
\end{table}

 For limb darkening we adopted a square root law that employs two coefficients, $x_{LD}$ and $y_{LD}$ per star and per passband. Note that  the standard \kepler transmission function in PHOEBE is the mean  over the 84 different channels of  the \kepler FoV.
  PHOEBE determines the coefficients by interpolating, for the  given atmospheric parameters,  its internal look-up tables.
 The gravity darkening and albedo coefficients were kept fixed at their theoretical values, $g = 1.0$ and $A = 1.0$ for radiative envelopes \citep{vonzeipel1924}. 

The primary effective temperature was kept fixed, as it is well known that the solution of a single light curve is sensitive only to relative values of the effective temperatures.  We used the value
$T_{\rm eff,1}= 6800$ K, which was provided by the analysis of the spectra. The uncertainties on the effective temperatures are essentially those from  spectra analysis, as the additional term  on $T_{\rm eff,2}$ from the fit is very small (a few degrees). A shift in $T_{\rm eff,1}$ reflects in a shift of $T_{\rm eff,2}$ in the same direction leaving the ratio almost constant.

 Figure~\ref{binlc} shows the fit and its residuals.  The larger scatter of residuals around minima  depends on the  smaller phase bins used for the eclipses. The remaining features around minima  are instead related to  the stellar surface representation in PHOEBE (and WD), in which the aligned  border of the first tile of each meridian produces  a seam along the central line facing the companion  (y=0, in the co-rotating frame centered on the star). This issue will be solved in the next PHOEBE version, still under development \footnote{see     http://phoebe-project.org/}.

\subsection{Combined solution}
  The system parameters from the light and radial velocity curve solution are collected in Table~\ref{sol}. 

  The system model corresponding to the best fit is formed by two similar stars: masses differ by 0.02  $M_{\odot}$, radii by 0.42  $ R_{\odot}$ and effective temperatures by less than 200 K. 
  
  There is an evident discrepancy between our combined solution and the results from the analysis of the disentangled spectra. According to  Table~\ref{atmos} the hottest star (by $\sim$ 70 K)  is the secondary component. On the other hand we know, as mentioned in the previous section, that the secondary, pulsating star is eclipsed at the widest and less deep minimum.  The shape of the light curve cannot be reproduced by a hotter secondary. Even considering the effect on the eclipse depth of the different  distance between the stars at minima because of eccentricity, stars of the same  temperature yield a secondary eclipse less deep  of only a few percent and the difference in depth will still decrease  (and then change of sign) for a hotter secondary, while the observed value amounts to about 10 \%.  A cooler secondary is, therefore, needed to reproduce the light curve shape.
  Considering that the line profiles of the secondary star are certainly affected by pulsations, we decided to trust and adopt in the following the $T_\mathrm{eff,2}$ value from the light curve modeling. 
  
  The uniqueness of the solution and the derivation of parameter uncertainties are discussed in the next section.

    \begin{figure*}[t!]
   \centering
      \includegraphics[width=8.5cm]{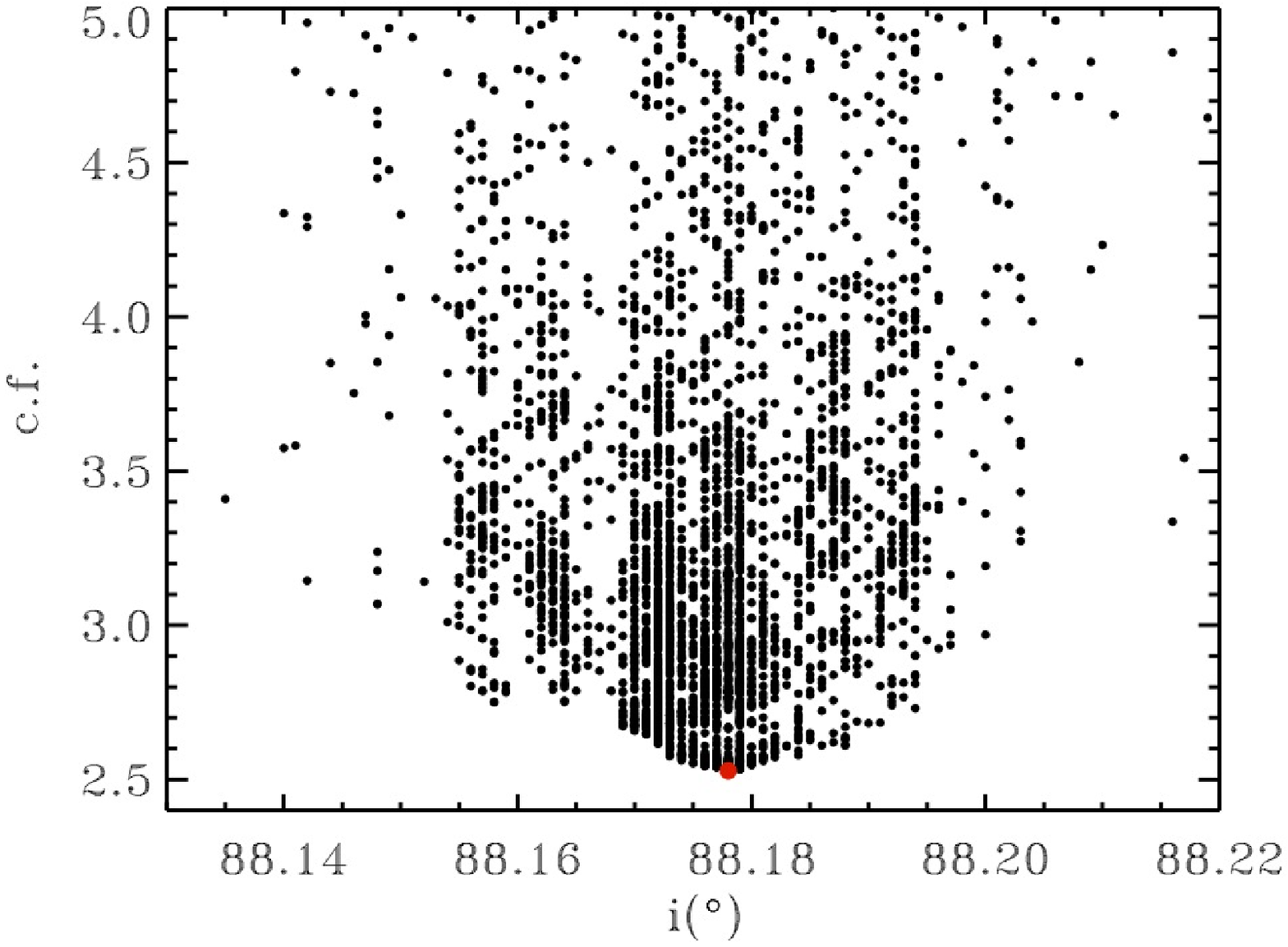}
   \includegraphics[width=8.5cm ]{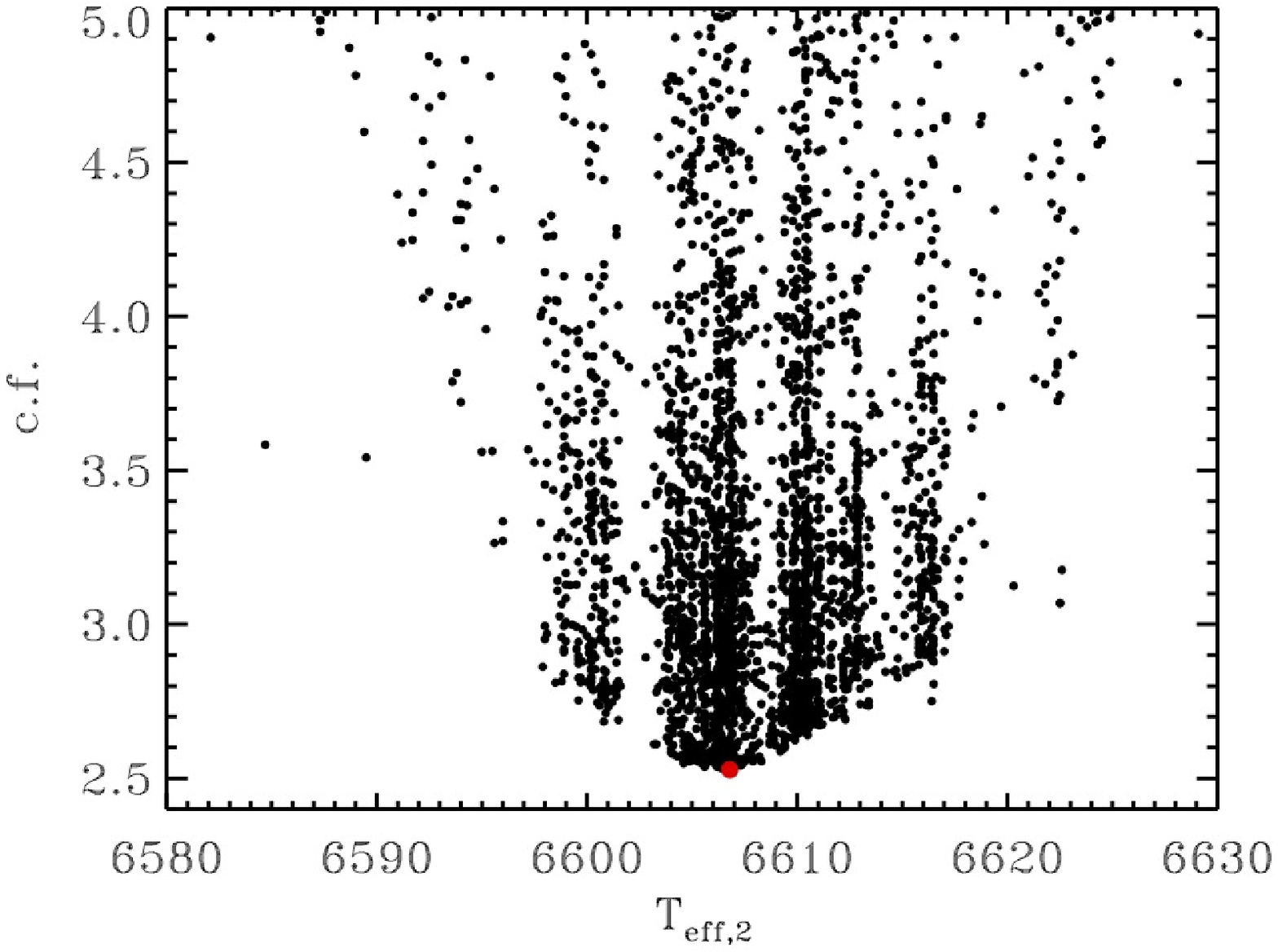}
   \includegraphics[width=8.5cm]{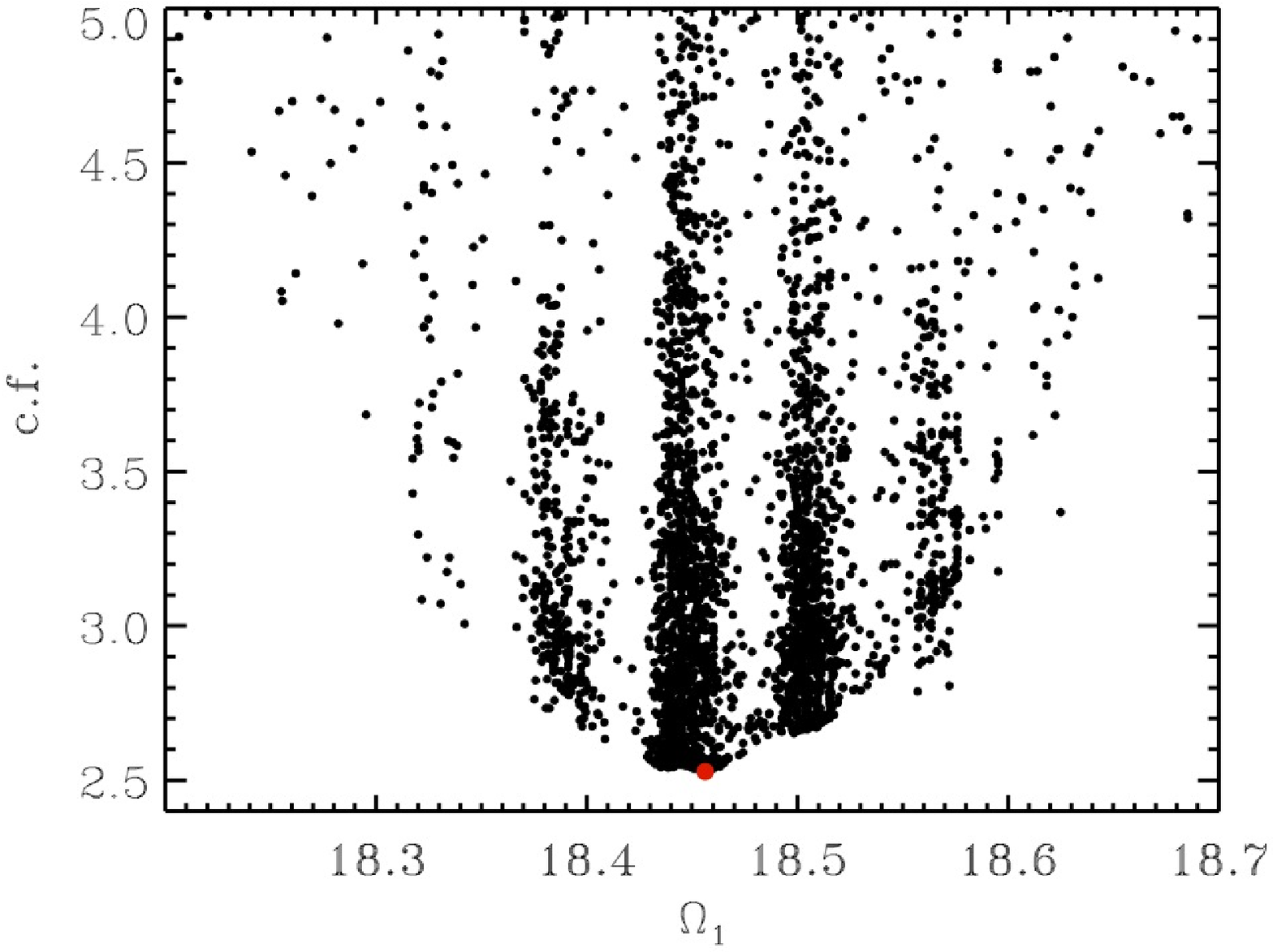}
   \includegraphics[width=8.5cm]{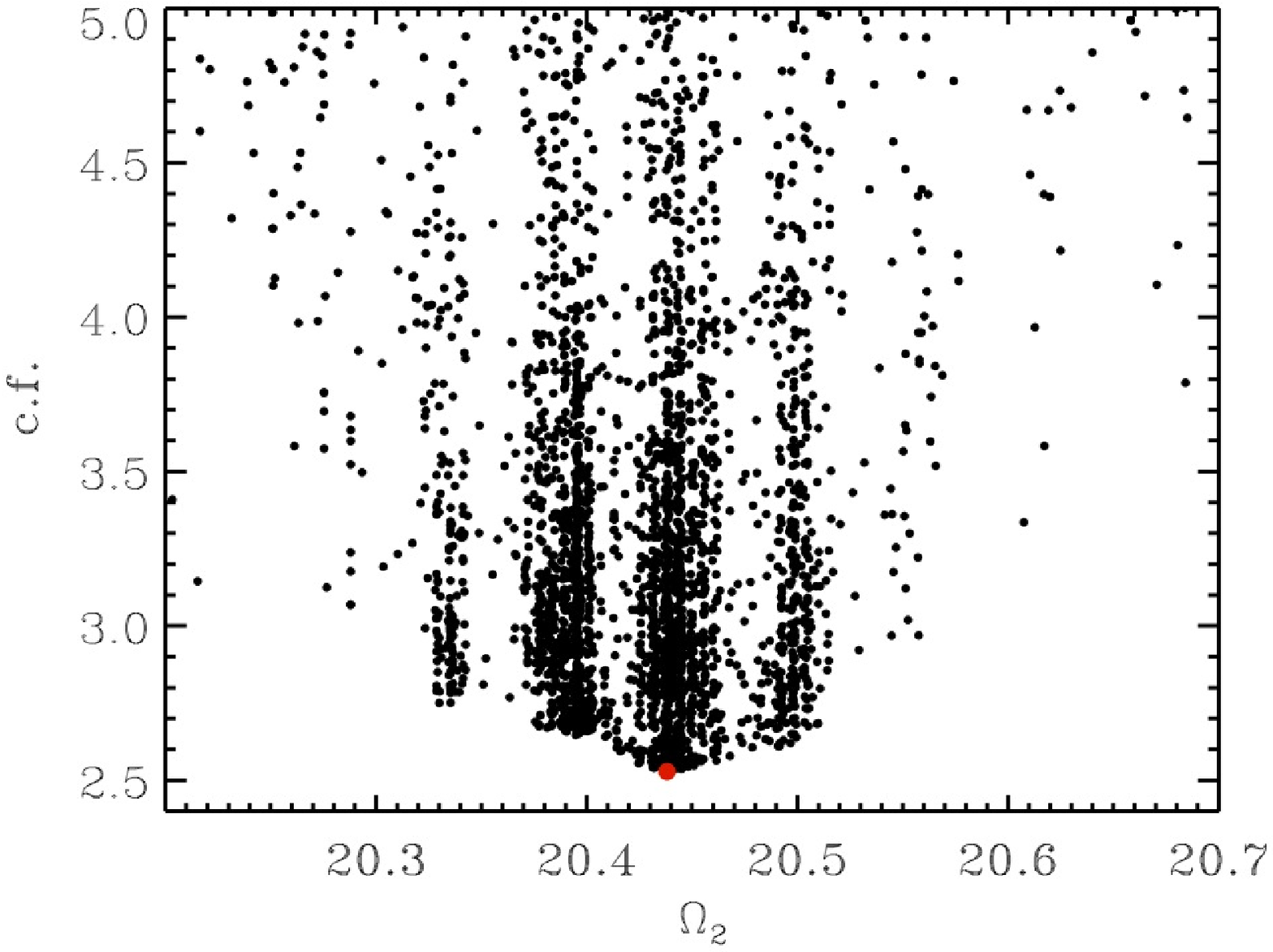}

         \caption{Blow up of the  PIKAIA results:   cost function value vs inclination (top left), secondary temperature (top right) and potentials (bottom panels). The minimum value is indicated by a gray (red in the electronic version) larger dot. }
         \label{pikscan}
   \end{figure*}

\subsection{Uniqueness of the solution and parameter uncertainties} 
\label{uniq}

The light curve fit is obtained  by minimizing  a cost function which measures  the deviation between model and observations in the space of adjusted parameters. In PHOEBE the minimization is done by differential correction, in the Levenberg-Marquart's variant, or by the Nelder and Mead's downhill simplex. As a consequence, it suffers from the well know problems of these methods: the minimization algorithm can be trapped in a local minimum or  degeneracy among parameters and data noise  can transform the minimum into a large and flat bottomed region,  or an elongated flat valley, rather than a single point. 
Besides, the correlation among the parameters implies uncertainties on the derived values which are significantly larger than the formal errors derived, for instance, by the least square minimization.  The disagreement between the analysis of photometry and spectroscopy was an additional reason to test the uniqueness of our photometric solution.

The search for the global minimum can be performed with a variety of techniques, which have  in common, however, the fact of being very demanding in terms of computing time.  \citet{prsazw05}, for instance, propose Heuristic Scanning and Parameter Kicking (a simplified  version of Simulated annealing). The former method was applied  to solve the CoRoT light curve of the eccentric binary HD~174884 \citep{cm09}.  

Another popular choice is that of genetic  algorithms, a class of optimization techniques inspired from biological evolution through natural selection of the fittest offsprings. A widely used implementation of genetic algorithms in astrophysical problems  is  PIKAIA\footnote{the code and related documentation are freely available from http://www.hao.ucar.edu/modeling/pikaia/pikaia.php} \citep{charb95}.  PIKAIA starts from a randomly chosen  population, whose individuals are characterized by the values of the parameters to optimize (the values are coded in their ``chromosomes''). Pairs of individuals are mixed through a breeding process yielding ``offsprings'',  the members of the new ``generation" are chosen on the base of their fitness, measured by a user-defined cost function.  Random mutation of some of the offsprings is also included.
 Thanks to the weighting of offsprings  with fitness and to the mechanism of breeding (which allows to incorporate information from the previous generation)  the search process of PIKAIA is much faster than in  random search algorithms (e.g., Monte Carlo).
   
We built an interface  program FITBINARY, which connects PIKAIA (v1.2) to the PHOEBE's underlying fortran code (a PHOEBE specific version of the  \citet{WD71}). We preferred to stick to the PHOEBE back-end WD-version because of  its many extensions,  and of its internal computations of outgoing fluxes and limb darkening in the \kepler bandpass. In this way, the FITBINARY results are fully compatible with PHOEBE modeling. 

The PIKAIA search   was performed on the same parameters adjusted in PHOEBE: inclination, surface potentials, and effective temperature of the secondary component, while the primary passband luminosity was recomputed at each step. We  made use of the creep mutation mode to jump over the so called Hamming walls (see the on-line PIKAIA documentation). The search was first conducted in  a large region around a provisional set of parameters,  to be  later on restricted to smaller intervals.   A typical scan (100 generation of 50 individuals) requires about 40 hours of computation on a 1.8~GHz Intel Core i7 processor with 4Gb RAM. For a non eccentric system the computation is much faster, since it is not necessary to recompute the surface geometry at each orbital phase.

Figure ~\ref{pikscan} displays the results of  the genetic algorithm, the four panels  display the  cost function (in arbitrary units) versus the four parameters. The clustering of points along vertical strips 
indicates the efficiency of the method in converging to the minimum.  The figure is actually a blow-up around the minimum, as the search was performed in parameter ranges larger than those displayed
($i=88.1- 88.3$, $\Omega_1=18 - 19$, $\Omega_2=20 - 21$, $T_\mathrm{eff,2} = 6550 - 6650$).  It is evident,  looking for instance at the $\Omega_2$ plot,  that a  ``blind" minimum search could  be trapped in any of the  parabolic shaped regions. 
The parameters of Table~\ref{sol} are derived from the parameters of PIKAIA's minimum. 

The uncertainties on the final parameters and the corresponding 1-sigma errors of Table \ref{sol}, were computed with Bootstrap Resampling (BR) both for the radial velocity and the light curve fit, 
according to the scheme described in detail in \citet{mr97} and \citet{cm09} and applied in MMG13. 
The main difference with the standard BR treatment  is that  the procedure is performed within the minimum already established by a single iterated solution (that is, using only one set of residuals and parameter derivatives computed in the global minimum). This is because the full computation of an iterated minimization for each bootstrap sample (we used 20000 samples)  is extremely  heavy in terms of computing time, especially in the case of an eccentric orbit, which requires re-computation of the binary surfaces  at each phase.
The Monte Carlo Markov Chain (MCMC) used in other cases (Da Silva et al. 2013, in preparation) is too demanding in terms of computing time in the case of an eccentric system.

\section{Pulsational properties} 
\label{puls}
\subsection{Frequency extraction}
\begin{figure}[ht!]
    \centering
    \includegraphics[width=9.cm,trim=30 10 0 0]{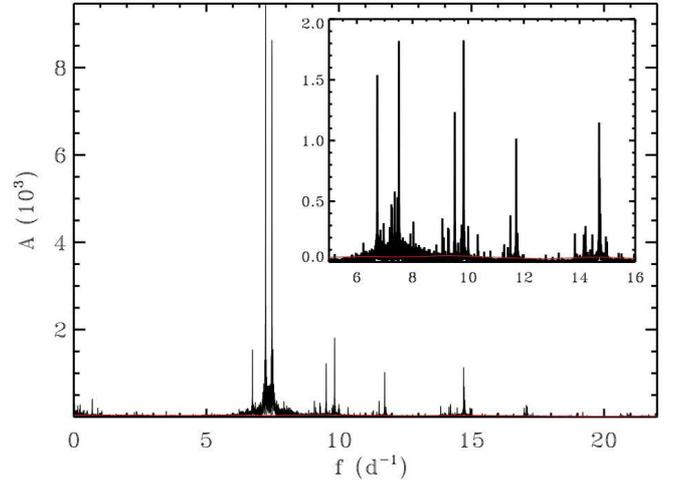}
 \caption{The final amplitude spectrum of \eb and, in the inset a blow-up of  the same spectrum after pre-whitening the two highest frequencies.  The gray  line (red in the on-line version)  is the significance threshold which  is four times the local noise (computed on the residual spectrum with a box of 3 \cd)}
        \label{frsp}
    \end{figure}

The  frequency spectrum, after subtraction of the final binary model, is shown in Fig.~\ref{frsp}.  The plot is restricted to the frequency interval containing meaningful features. Since the residuals of Fig.~\ref{binlc} are  dominated by  deviations around minima, we computed the frequency spectra both for the full residuals  and for the out-of-eclipse  sections. The spectra are very similar, the main difference is the presence, in the first case,  around the two dominant frequencies of a large number of  low-amplitude peaks which are spaced by multiples of the orbital frequency. The spurious nature of these features is evidenced by comparison with the out-of-eclipse Fourier spectrum, in which  they disappear. 
 For this reason we adopt and discuss  in the following  the  results obtained from the light curve cleaned of the EB signal but excluding the eclipse sections.
    
   As in our preliminary analysis we used both SIGSPEC and PERIOD04 for frequency extraction.  SIGSPEC has the advantage of automatically computing the frequency components and   their spectral significance, $\xi$, which  is related to the false alarm probability of detecting as a peak a feature actually due to (white) noise: $\xi= -\log( \Phi_{FA} )$  \citep{sigspec}.   PERIOD04 was used to  check the results,  to compute the local noise spectrum,  and to derive the  S/N ratio.
    
A commonly adopted frequency significance  threshold  is that suggested by  \cite{breger93}, i.e.   a S/N value of the amplitude  $\ge$ 4. This value, however, was actually derived for ground-based multisite campaigns and  is also based on the assumption of random noise. This may not be the case for  \kepler data  or pulsating stars which can show significant red noise \citep{kallinger10}.
According to \citet{sigspec2} the approximate correspondence between spectral significance and S/N ratio yields $\xi(S/N=4)=5.47$. We checked, however, that in  our dataset  we should not go below   $\xi\simeq12$ to have a S/N>4 everywhere (and in particular at low frequencies). We adopted therefore this more conservative criterion for significance.

 Table \ref{freqs} lists the first sixty out of about four hundred frequencies in order of detection and their spectral significance, the full table is available as on-line material. 
  The uncertainties in the table have  been computed according to \cite{kallinger08} and are sensibly larger than the formal errors of the LS fit.  The obvious combinations of frequencies and the overtones of the orbital frequency ( $f_{ \rm{orb}}=0.0385327$) are also listed in the comments. 
 
  The inspection of Fig.~\ref{frsp} and Table~\ref{freqs} reveals that  pulsations occur with frequencies in two separate regions of the  spectrum: a high-frequency region above 6 \cd, containing most of the signal, and a low-frequency one (around or below 1 \cd), whose strongest frequency, $f_{13}$ of the table, has an amplitude $\sim$4 \% of $f_{1}$.
 The first region is where one expects p-mode pulsations for classical $\delta$~Sct stars, the second is the typical region of g-modes in $\gamma$ Dor stars. 
 
 As our pulsator is a highly eccentric binary one can also expect that the orbital motion and the varying tidal forces along the orbit  affect  the pulsations.
 According to \citet{shibahashi12}, the Doppler effect due to the orbital motion  of a pulsating star in a binary causes a phase modulation of pulsation. That   yields,  in the Fourier   decomposition of the time series, a multiplet  with the original frequencies  split  in components separated by the orbital frequency. The amplitude of the multiplet components depends on the binary parameters and on the ratio of the orbital and pulsation periods.  Using the relations of the above mentioned paper, we derived the ratio of the  amplitude of the first multiplet components ($\pm i f_{ \rm{orb}}; i=1,2$) to the central  frequency  for the case of an eccentric binary (note that symmetric $\pm$peaks have the same amplitude).   We obtain that the amplitude of this first component should be $\sim$1.5\% of the central frequency,  while  that of the second 0.3 \%;  the latter is  a too small value for our detection threshold, even for the strongest frequencies. 
 In our frequency list we find at least the first component ($i=1$) of  the multiplets around $f_{1, 2}$, with an amplitude somewhat  higher than expected, but at least of the right order
 ($f_{4}, f_{19}, f_{45}$, of Table~\ref{freqs}, while   $f_{81}=f_2-f_{{\rm orb}}$ can be found in the complete on-line version).

 We could not find, instead,  a convincing evidence on pulsation  of the star rotation (which corresponds to  a rotation frequency  $f_{\rm{rot},2}=0.167$ \cd, according to the measured \vsin and the derived values of inclination and stellar radius). The  ratio between the rotation and break-up velocities is small enough  ($v_{\rm rot,2}/v_{\rm b,2} \sim 0.008$)  to expect equidistant multiplets \citep{reese06}, which are easier to spot, but  the high number of frequencies and the uncertainty on the rotation velocity  prevent detection. 

\subsection{High frequency variability}
 An estimate of the pulsation frequency regime for a $\delta$~Sct can be derived by the relation between the frequency of a radial mode and the mean density through the  the  pulsation constant $Q$
 \citep{stell79}:
 \begin{equation}
 \label{qconst}
 Q= P_{\mathrm{osc}} \sqrt{ \frac{\overline{\rho}}{\rho_{\odot}}}. 
 \end{equation}
 Typical $Q$ values in the $\delta$~Sct domain are  $Q=0.033$ for the fundamental radial mode and  decrease    to 0.017 for the  third overtone.
  For single stars the mean density is inferred from $T_{\mathrm{eff}}$, the absolute magnitude,  and surface gravity, but in our case we have an accurate determination of mass and radius from the binary model.
 The pulsation frequencies corresponding to the fundamental and the first three overtones for the secondary star, with the  mentioned $Q$ values,  are  in the range 7.8 - 15 \cd, for the primary star the interval is from 6.5 to 12.5 \cd.  That corresponds  to the frequency  interval where we find most of the signal. A more precise computation of the excited frequency ranges, based on theoretical models,  is presented in Sect.~\ref{abspar}.
  \begin{figure}[htb]
    \centering
    \includegraphics[width=9.cm,trim=30 10 0 0]{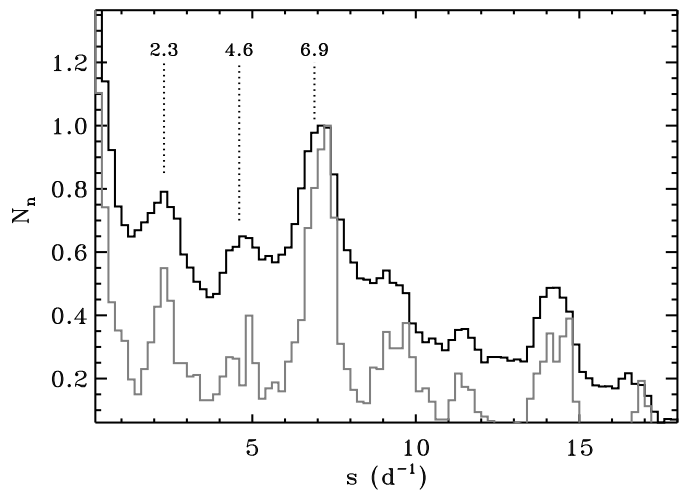}
 \caption{The distribution of differences among the detected frequencies (every frequency was subtracted from all the others and the plot is restricted to the positive values). For easier comparison each histogram is normalized to the local maximum around 7 \cd. The black line refers to the whole set of 403 frequencies, the gray line to those with amplitude larger than $10^{-4}$.  The histograms show that there is a preferred spacing  around 2.3 \cd which is clearly visible in both distributions. According to \citep{breger09} the spacing corresponds to that between radial modes. The multiples of the spacing are also indicated by dotted lines. }
        \label{hist_frdiff}
    \end{figure}

As \citet{breger09} pointed out,   the numerous observed frequencies of $\delta$~Sct  often appear in groups which cluster around the frequencies of radial modes and show some regularity in spacing. 
The fact is put in relation with mode trapping in  the envelope  which   increases the probability of  detection \citep{dziemb90}. These modes are the non-radial counterparts of the acoustic radial modes and, for low $\ell$ degree, have frequencies close to those of radial modes.  In the case of $\delta$~Sct pulsators theory predicts a much higher efficiency of mode trapping for $\ell=1$ modes, see also \citet{breger09}.

The above-mentioned authors introduce a theoretical diagram displaying the average separation of the radial frequencies versus the frequency  of the lowest-frequency  unstable radial mode (their Fig.~8, $s-f$ diagram).  The values are arranged on a grid with the different radial modes and surface gravity as parameters. The location of  a pulsator in that diagram allows, therefore, to estimate its surface gravity (if the radial mode is assumed or known).  The histogram of the frequency difference  is shown in Fig.~\ref{hist_frdiff} both for the whole set of  frequencies and for those with an amplitude $A > 10^{-4}$ ($\sim 100$ frequencies). There is an evident  pattern, indicating a separation of the radial modes of $s \simeq 2.3$ \cd. One sees in fact a first peak at 2.3 \cd and others at approximately two and three times the value  (the separation of radial modes is not exactly constant  with increasing order). 
As we already have a precise estimate of the gravities, we used the value of $s$ to derive the frequency of the fundamental radial mode of the secondary, which is $\sim 7.4$ \cd, i.e. very close to  the values of $f_{2}$.   We know that both $f_1$ and $f_2$ belong to the secondary component, if - as plausible - we  assume that $f_2$ is its fundamental radial mode,  $f_1$ should be a non radial one. The lower frequencies in the same domain  could be either  non-radial modes of  the secondary or radial/non-radial modes of the primary component.

\subsection{Low frequency variability}

 The low frequencies detected in the range of high order g-modes, typical of   the $\gamma$~Dor variables,  could  indicate a hybrid nature of the $\delta$~Sct pulsator.
 
 According  to the asymptotic approximation \citep{tass80}, the angular frequency, $\sigma_{n\ell}$,  of high order g-modes of radial order $n$ and angular degree $\ell$ propagating between two convective regions is given by:
\begin{equation}
\label{tassoul}
\sigma_{n\ell}=\frac{\sqrt{\ell (\ell+1)}}{\pi (n+1/2)} \Im
\end{equation}

\noindent where $\Im=\int_{r_1}^{r_2} \frac{N}{r} dr$ is the integral of the Brunt-V\"ais\"al\"a frequency ($N$), weighted by the radius, along the  propagation cavity.

For the stellar structure and frequencies of interest  here $r_1$ and $r_2$,  and hence $\Im$, are weakly dependent on the mode. As a consequence, the first order asymptotic approximation also predicts that high order g-modes should be regularly spaced in period, with a mean period spacing:
\begin{equation} 
\label{pspac}
\Delta P_\ell=\frac{2 \pi^2}{\:\Im\:\sqrt{\ell (\ell+1)}}.
 \end{equation}

\begin{figure}[t!]
    \centering
    \includegraphics[width=9.cm]{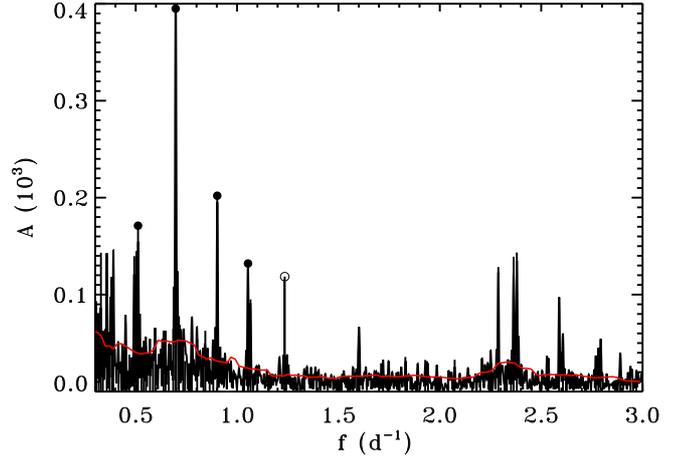}
 \caption{The frequency spectrum in the $\gamma$~Dor domain after pre-whitening with the first  twelve frequencies  of Table \ref{freqs}. The highest amplitude peak corresponds to $f_{13}\simeq18 f_{\rm orb}$. The gray (red) line is the estimate of local noise by PERIOD04  (with a box of 0.2 \cd). The black dots denote the values used to derive the period spacing. The open circle denotes the frequency $f_{193}=1.2350$ \cd, which according to that period spacing value has $n_{f_{193}}=39$. Note that this last frequency is also very close to 32$f_{\rm orb}$. }
        \label{frgd}
    \end{figure}

\citet[][hereafter M05]{moya05} derive from Eq.~\ref{tassoul}  a straightforward  relation between radial orders,   frequencies,  and corresponding ${\Im}$ values, for modes with the same $\ell$ value:
 \begin{equation}
 \label{freqratio}
 \frac{f_i}{f_j}\approx \frac{n_j+1/2}{n_i+1/2} \: \frac{{\Im}_i}{{\Im}_j}\approx \frac{n_j+1/2}{n_i+1/2};
 \end{equation}
     
the second approximation is justified by the fact that, for the modes of interest here, $\Im$ is very weakly dependent on the order.
 We checked  if such a relation was verified among the frequencies  in the  interval 0.3 - 3.0 \cd. We computed the expected $n$ ratios for radial orders $1<n<120$, in the hypothesis of same degree $\ell$, and compared the results with the observed ratios of the four locally highest amplitude  frequencies  ($f_{13}$, $ f_{41}$, $f_{47}$, $f_{55}$, see Fig.\ref{frgd}), excluding from the search the obvious  combinations, such as $f_{61}=0.4930$~\cd $=f_4-f_1$ . We assumed a  tolerance for each ratio of 0.004 \cd, which takes into account both the deviations of the asymptotic relation with respect to the theoretical models, see M05,  and the uncertainties on the derived frequencies.  We found, indeed, a few combinations of radial orders reproducing,  within the uncertainties, all the observed ratios (Table~\ref{frratio}).

 Each combination allows to estimate from Eq.~\ref{tassoul}, if a value of $\ell$ is assumed, the mean value of ${\Im}$,  i.e. a quantity which can be readily compared with theoretical models.
Cancellation  due to surface integration of the observed flux suggests low $\ell$ values and, in
 our case, a clue to infer the $\ell$ value is offered by the fact that the highest amplitude frequency among the four selected  is very close to an overtone of the orbital frequency,  $f_{ 13} \simeq18 f_{ \rm{orb}}$.  Since tidal action is more efficient in exciting $\ell=2$ modes \citep[e.g.,][]{koso83},  $\ell=2$ is the straightforward assumption.  Table~\ref{frratio} lists  the values of the integral of the Brunt-V\"{a}is\"{a}l\"{a} frequency, $\Im $ and of its inverse in seconds derived under this hypothesis. The comparison with the theoretical models is presented in the next Section. The lines in boldface highlight the combinations in better agreement with the results of that section.

The mean period spacing  (and its standard deviation) corresponding to the two $n$ series highlighted in  Table~\ref{frratio} (and $\ell=2$) is the same: $\Delta P= 1736^{\rm s} \pm 19^{\rm s}$.

\begin{table}[t!]
\caption{The combinations of radial orders, $n_i$, following the relation in Eq.~\ref{freqratio} for $ f_{47}$,  $f_{13}$, $f_{41}$, and $f_{55}$ of Table~\ref{freqs}, the corresponding integral of the
the Brunt-V\"{a}is\"{a}l\"{a} frequency  for $\ell=2$ (mean value $\pm$ its standard deviation, and its inverse).}    
\label{frratio}      
\centering
\begin{tabular}{cccccc}
\hline\hline
$ n_{f_{47}}$ & $n_{f_{13}}$ & $n_{f_{41}}$ & $n_{f_{55}}$ &  $ \Im (\mu$Hz) & $ \Im$$^{-1}$(s) \\ 
\hline
 48  &  35  &  27  &  23 &  367.8 $\pm$ 0.4  &  2718 $\pm$ 3 \\
 60  &  44  &  34  &  29 &  460.8 $\pm$ 1.2  &  2170 $\pm$ 6\\
 
 83  &  61  &  47  &  40 &  635.0 $\pm$ 1.4  &  1575 $\pm$ 4\\
 85  &  62  &  48  &  41 &  648.6 $\pm$ 1.3  &  1542 $\pm$ 3\\ 
\bf 96  & \bf 70  &\bf  54  & \bf 46 & \bf 729.8 $\pm$ 2.1  & \bf 1370 $\pm$ 4\\
\bf 97  &\bf  71  &\bf  55  &\bf  47 & \bf 741.6 $\pm$ 1.8  & \bf 1348 $\pm$ 4\\
108  &  79  &  61  &  52 &  822.8 $\pm$ 1.0  &  1215 $\pm$ 2\\
109  &  80  &  62  &  53 &  834.5 $\pm$ 2.9  &  1198 $\pm$ 4\\
112  &  82  &  63  &  54 &  852.7 $\pm$ 1.7  &  1173 $\pm$ 2\\
120  &  88  &  68  &  58 &  915.7 $\pm$ 1.1  &  1092 $\pm$ 1\\
\hline
\end{tabular}
\end{table}
 The maximum number of frequencies whose ratios follow the relation  in Eq. \ref{freqratio}  is $N_m=4$. We tested that by adding to the  quadruplet the other clearly detected frequency in the interval  $f=0.5-2.5$ \cd and, within the tolerance given above, we found no consistent solution. On the other hand, by adopting the derived period spacing,  we found a few frequencies which are close to the expected values for some $n$,  for example  $f_{193}=1.2350$ \cd ~($n_{f_{193}}=39$), which is also close to an overtone of the orbital period ( 32$f_{\rm orb}$). However, given the increasing uncertainties on lower amplitude frequencies and the intrinsic uncertainties in the procedure using asymptotic relations,   we prefer to restrict our analysis to the  firmly established quadruplet.

      \begin{figure*}[t!]
     \centering
     \includegraphics[width=8.7cm,]{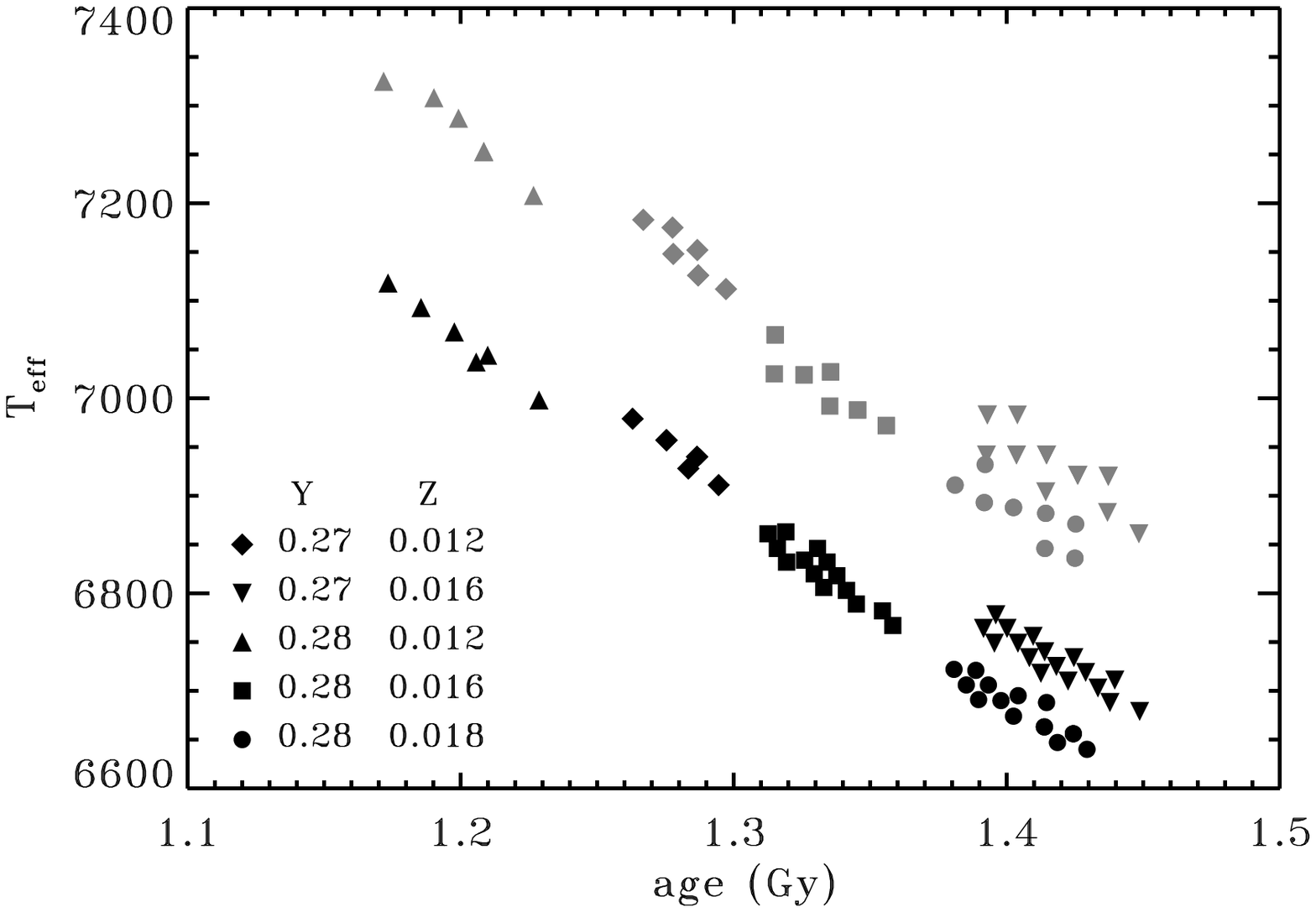}
     \includegraphics[width=8.7cm]{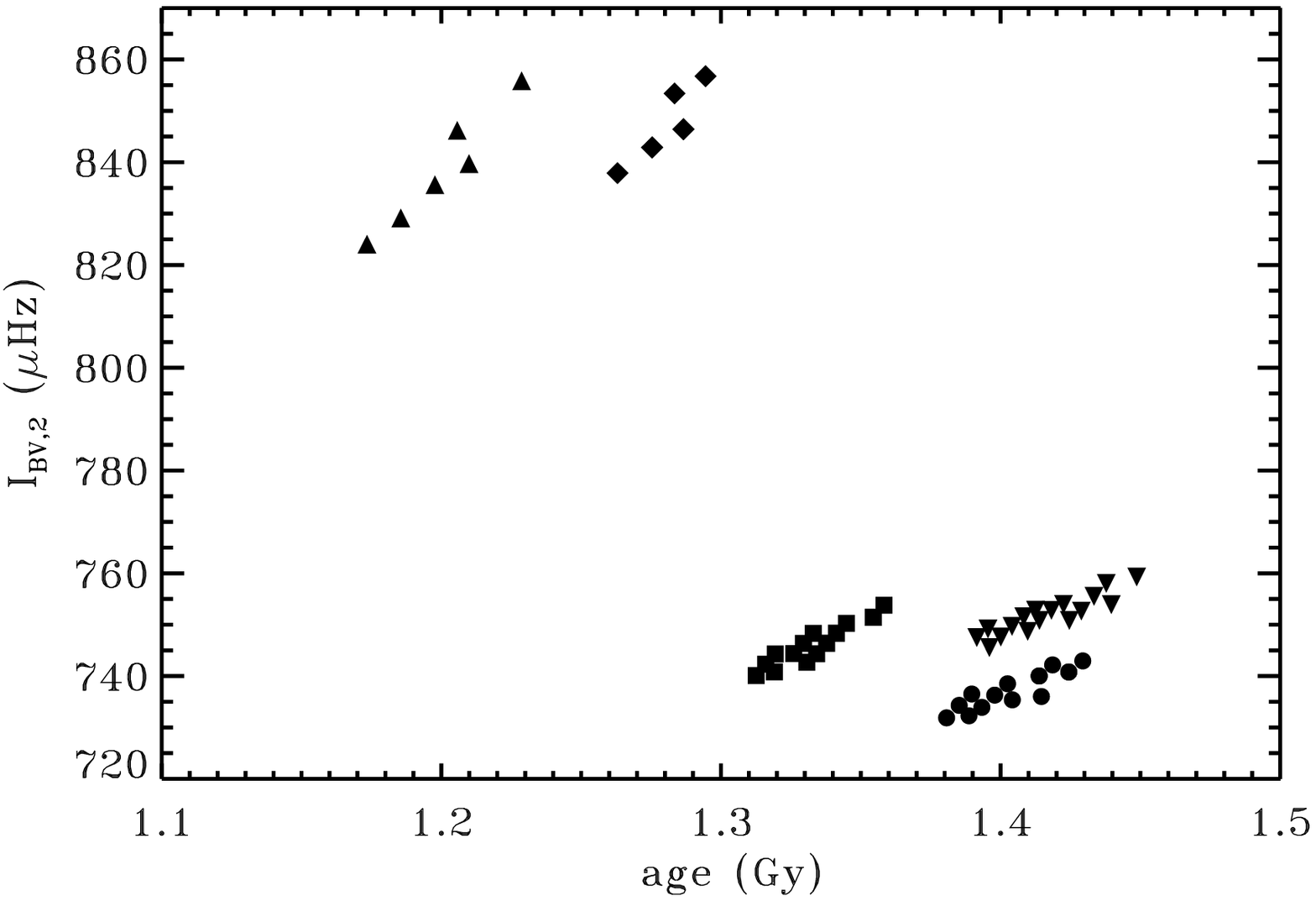}
     \includegraphics[width=8.7cm]{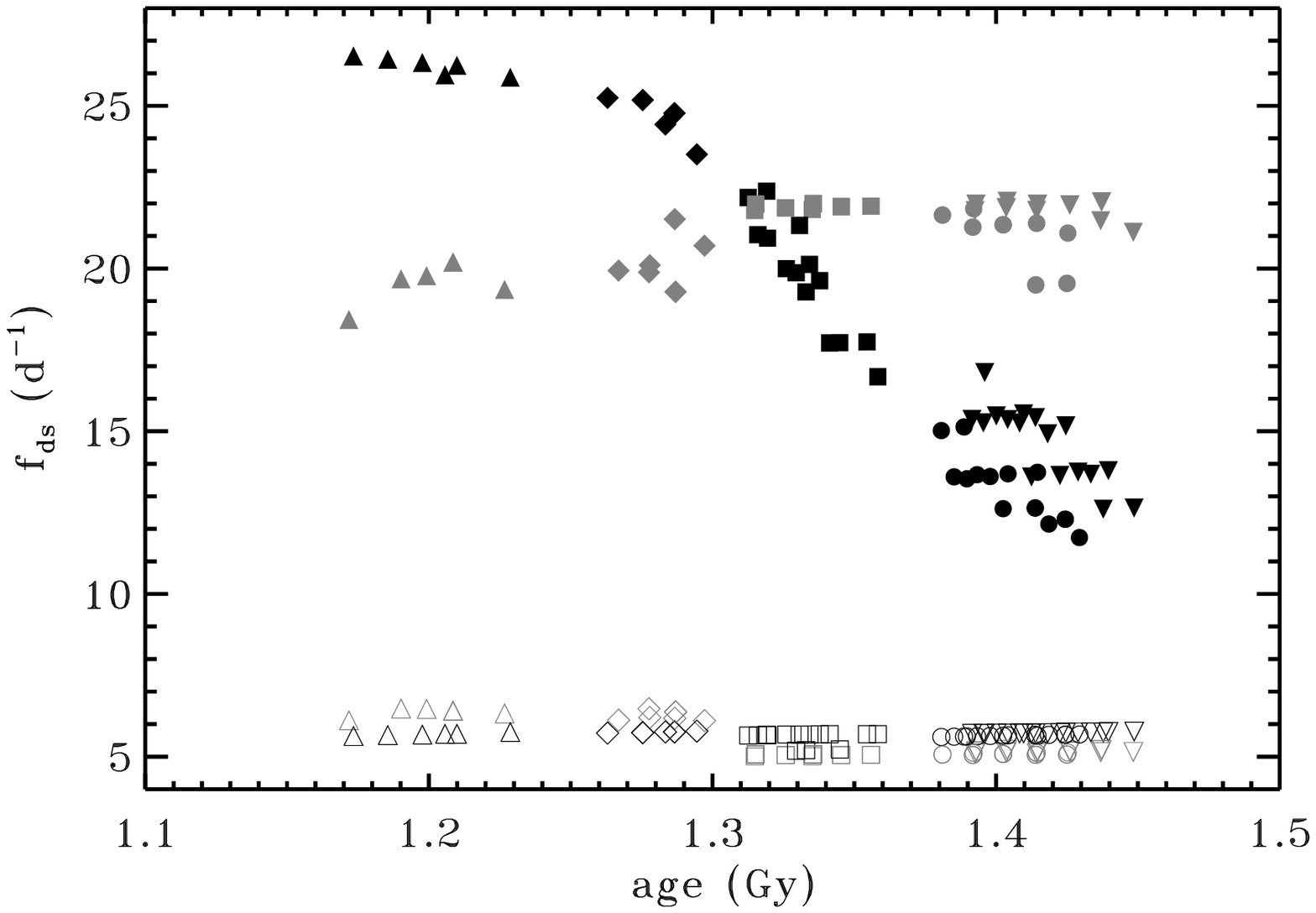}
     \includegraphics[width=8.7cm]{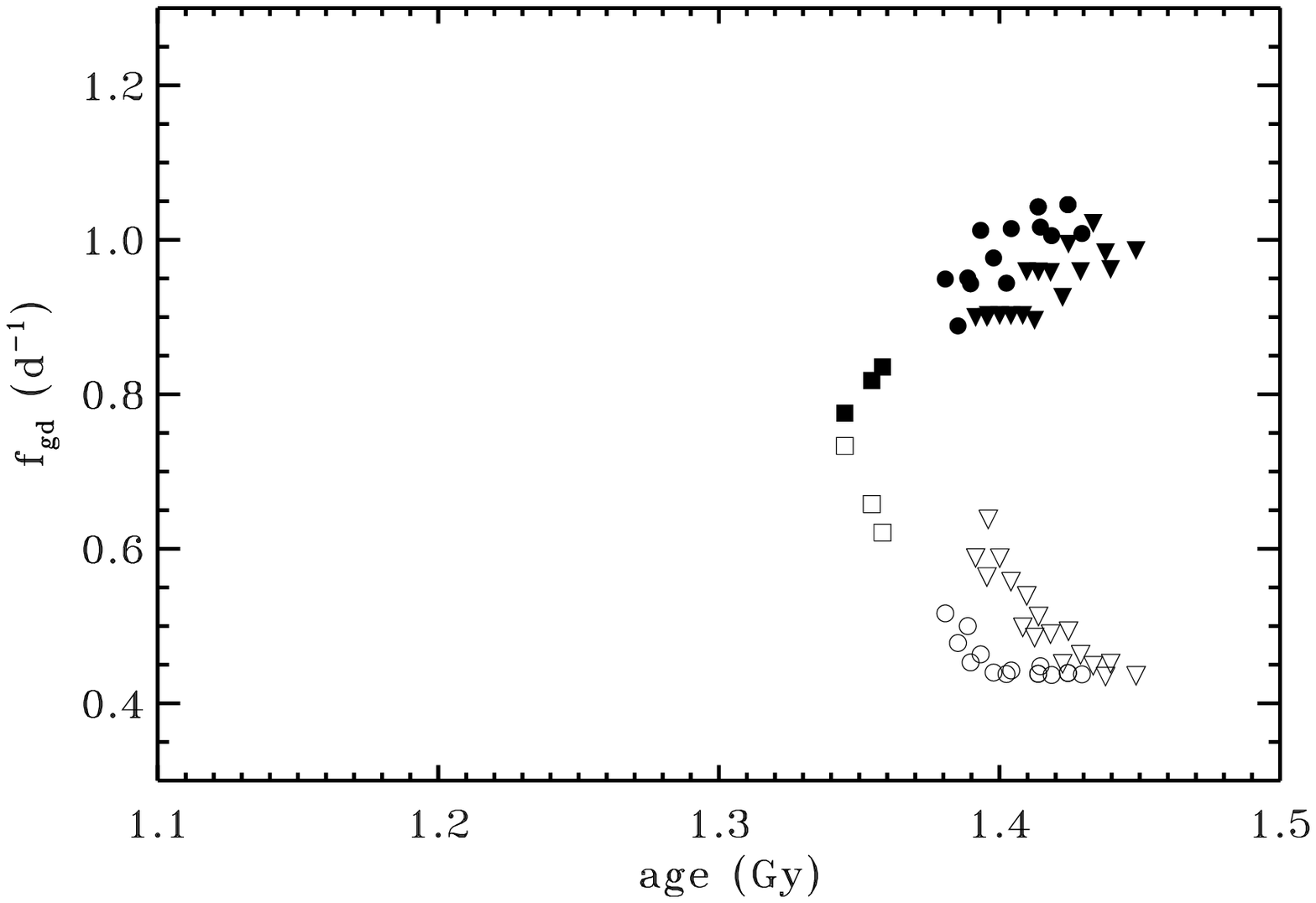}
     \caption{The properties of the stellar models selected as explained in the text. In all panels different quantities are plotted as function of the model age, the symbol shape depends on the chemical composition, the grey symbols correspond to the primary star, the black ones to the secondary.   Top left panel:  T$_{{\rm eff}}$ for both stars;  top-right panel: the integral of the Brunt-V\"{a}is\"{a}l\"{a} frequency for the secondary component; bottom left panel: the minimum and maximum excited frequency in the $\delta$~Sct domain for both stars  (empty and full symbols denote, respectively,  the minimum and maximum values), bottom right panel: the  minimum and maximum excited frequency in the $\gamma$~Dor domain for the secondary.}
       \label{modelplot}
     \end{figure*}
\section{Physical properties of \eb} 
\label{abspar}
Further information on  the physical properties of the system can be derived by comparison with theoretical models. As in previous studies \citep{cm13}, we use the evolution of the stellar radii with time, combined to the constraint of coevality, to infer the system age. 
The stellar radius evolution  instead of, for instance,  the position in the H-R diagram, is preferred because the radii are directly  and precisely determined from the radial velocity and light curve  solutions, while the computation of temperature and luminosity  implies the use of color transformations and bolometric corrections.

The evolutionary tracks were obtained from stellar evolution modeling with the code CLES  \citep[Code Li\'{e}geois  d'\'{E}volution Stellaire,][]{scuetal08} with the following micro-physics:  equation of state from OPAL  \citep[OPAL05,][]{ronay02};   opacity tables, for two different solar mixtures \citep[][GN93 and AGS05]{gn93, 2005ASPC..336...25A},  from OPAL  as well \citep{iglrog96},  and  extended to low temperature with the \citet{fergal05} opacity values;  and nuclear reaction rates from the NACRE compilation \citep{angulo99}, except for  $\mathrm{ ^{14}N(p,\gamma)^{15}O}$, updated by \citet{formi04}. 

Convection is treated by the mixing length theory \citep{bv58}, with the  value of $\alpha_{\rm MLT}$ fixed to 1.8 (at any rate, in the temperature range of interest here, its value does not affect the stellar radius). Models with and without overshooting were computed with the extra-mixing length expressed in terms of the local pressure scale height ($H_{\rm p}$), as $\alpha_{\rm OV}$, and  $\alpha_{\rm OV}$ values equal to 0.15 and 0.20.

We have computed evolutionary models for masses between 1.82 and 1.91~\ms, with a step of 0.005~\ms. For the chemical composition we assumed different values of the metal mass fraction ($Z$), from 0.007 to 0.018, and two values of the helium mass fraction $Y=0.27$, and 0.28.   
Because of the abundance anomalies the value of  Z  is   not well constrained. As mentioned in  Sect.~\ref{spectr_anal}   the primary shows some features of Am-Fm stars, except for the  Fe overabundance. The Am-Fm phenomenon disappears or decreases with  surface gravity \citep{kunzli98}. Therefore, the slightly evolved status of the primary ($\log g=3.6$) could explain the particular chemical pattern.
It is widely accepted that the Am-Fm phenomenon is the consequence of microscopic diffusion and radiative acceleration, together with macroscopic processes such as rotational induced mixing or stellar wind. Radiative accelerations are not implemented in CLES,  hence we cannot fit the chemical pattern at the surface. On the other hand, if photospheric abundances are the results of selective diffusion processes, we have no information on the chemical composition of the whole star. Because of the above mentioned anomalies the Fe abundance  cannot be used as a proxy for [M/H] to be used in stellar modeling.

To derive the age of the system, we have selected the models such that primary and secondary have at the same age the corresponding masses and radii, and have a difference of effective temperature of  of $\sim$200~K, as suggested from the light curve solution. Given the similar mass of the two components, it is  possible to fit the difference in radii only if  the two stars are in a different evolutionary  stage, and we could get this result only if overshooting is  included in the computations. We have then considered solutions in which both stars have the same extra-mixing in the central regions ($\alpha_{\rm OV}=0.15, 0.20$), and  also the possibility of a different efficiency of  the central mixing process in either component, by using  $\alpha_{\rm OV}=0.20$ for the primary and $\alpha_{\rm OV}=0.15$ for the secondary, and viceversa.

In the following we present and discuss the models relative to the AGS05 solar mixture, which provide similar results to those with GN93 and  was used as reference in Table~\ref{chemcomp}. 

The constraint on the effective temperature of the components rules out models with the lower metallicity values, the best match is obtained for $Z=0.016$ and 0.018 which means $[M/H]~+0.1$ or +0.15 depending if we assume AGS09 \citep{asplund2009} or AGS05, respectively.  If,  in addition,  we fix the primary temperature to $6810 \pm 140$ (i.e. $\pm 2 \sigma$),  as obtained  from spectroscopy,
we can discard all models with   $\alpha_{\rm OV}=0.15$ in both stars, and almost all those with different $\alpha_{\rm OV}$ in the two components (with the exception of three models with $\alpha_{\rm OV,1}=0.15$  and $\alpha_{\rm OV,2}=0.20$). The rejected models, in fact,   reproduce  the temperature difference but provide a too high value of T$_{{\rm eff,1}}$. 

The constraints listed above select a sample of 59 stellar models with equal $\alpha_{\rm OV}$ and different chemical abundances (we dropped the three more models with different
$\alpha_{\rm OV}$ after checking that the results do not change).  
The system age of the whole sample varies  from   1.17 to 1.49 Gy  (see the top-left panel of Fig.~\ref{modelplot}) and   is significantly reduced, to to 1.38 -- 1.49 Gy, fixing the primary temperature. 
The other panels of the figure show the value of  $\Im$  and the maximum and minimum excited frequencies in the $\delta$~Sct and $\gamma$~Dor domains, as computed with the non-adiabatic pulsation code MAD by \citet{MAD03}.
From  inspection of these panels we can conclude that  the derived age range corresponds to: 
\begin{itemize}
\item a value of the Brunt-V\"{a}is\"{a}l\"{a} for the secondary component between 730 and 760 $\mu$Hz (top-right panel), note that the same quantity for the primary is $\sim 1100$~$\mu$Hz,
\item a minimum and maximum value of the excited frequencies, in the $\delta$~Sct domain, of 5--17 \cd for the secondary and of 5--22 \cd for the primary (bottom left panel),
\item a minimum and maximum value of the excited frequencies, in the $\gamma$~Dor domain, of 0.4--1.1 \cd (bottom right panel), the plot is for  secondary only as the excitation mechanisms is not at work in the primary.
\end{itemize}
All these values are quite consistent with what was obtained in the previous Section. 
Both stars can pulsate in the $\delta$~Sct domain (but our computations do not provide the amplitude of the frequency components so we cannot tell which one is actually detectable), while the driving mechanisms for $\gamma$~Dor pulsations can work in the secondary only, which could be therefore a genuine $\delta$~Sct/$\gamma$~Dor hybrid pulsator. On the other hand our computations do not take into account the presence of an ``external'' excitation mechanisms, as is the varying tidal force in an elliptic orbit, and the presence of an orbital frequency overtone among the frequencies  detected in the $\gamma$~Dor domain could be  a clue about such an origin.
To discriminate  between the two hypotheses, we computed from Eq.~\ref{tassoul} the frequencies corresponding to $\ell=1$ modes and to the selected $\Im$ values. The underlying idea was that, at variance with tidal forces,  an intrinsic excitation mechanism would work also for $\ell=1$ modes. We  found  that for $\Im=729.8$ $\mu$Hz, three of the detected frequencies ($f_{68}=0.3573$, $f_{100}=0.3284$, and $f_{279}=0.4686$ \cd)  correspond, within their errors, to expected $\ell=1$ values. As, however, the  theoretically expected g-mode frequencies are closer and closer for decreasing frequency values, an occurrence due to pure coincidence is also possible. 

\begin{figure}
     \centering
     \includegraphics[width=8.7cm,]{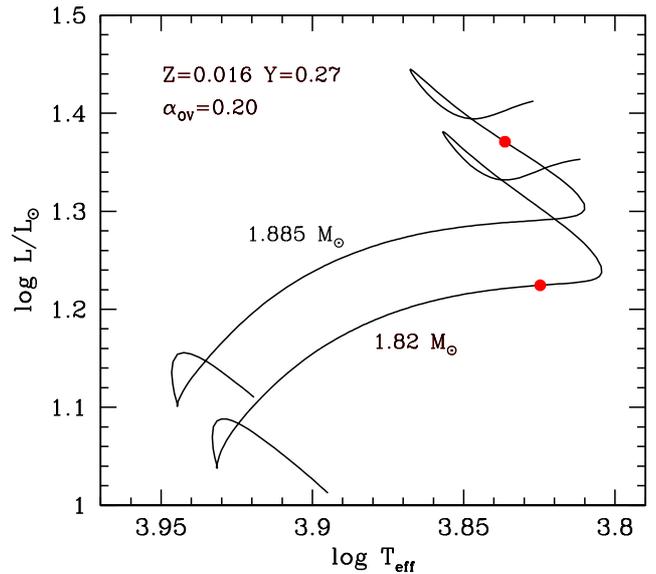}
      \caption{An example of  evolutionary tracks for a pair of stellar models fulfilling the constraints explained in the text, and with Y=0.27, Z=0.016 and  $\alpha_{\rm OV}=0.2$. For this  particular pair  $T_{\rm eff,1}=6860$~K  and $T_{\rm eff,2}=6680$ K, the requested difference  in radius with similar masses is fulfilled thanks to  a more evolved stage of the primary component.}
       \label{hrd}
 
     \end{figure}

In Fig.~ \ref{hrd} we show a representative pair of evolutionary tracks  from the selected sample (that in best agreement with all the constraints). To be noticed that the  fundamental radial mode of the secondary model is, in this case, $f\simeq 7.5$ \cd, i.e. in excellent agreement with the finding from frequency differences of the previous Section.
\section{Discussion and conclusions}
Thanks to the unprecedented quality of the Kepler data and to a long spectroscopic follow-up involving several different instruments we have been able to achieve a comprehensive description of this challenging binary. As it frequently happens, the availability of higher quality data is at the same time a source of achievements  but as well of new challenges. 

The \kepler light curve was intractable with the  method currently used to analyze pulsating star in binaries \citep[e.g., MMG13 ][]{jonas13}, given the high eccentricity and the deep eclipses, and hence the  modified  oscillatory pattern during the deep eclipse of the pulsator. 
To face this problem we developed a new procedure to extract the orbital (i.e. binary only) contribution to the light curve, which takes into account the effect of eccentricity and (at least at first order) of the eclipse of the pulsating star.  The straightforward development would be a detailed detailed analysis  of the eclipse of the pulsator with the  eclipse mapping technique \citep{biro11}, which would also allow to extract  information on the observed modes.

The partial disagreement between the parameters derived from photometry and spectra analysis was a motivation for an extended scan of the parameter space of the light/RV curve fits, to rule out the possibility of a wrong solution corresponding to  a secondary minimum. To this purpose  a genetic minimization algorithm was implemented in combination with PHOEBE.

The final model of \eb depicts a binary with components of similar mass and temperature but different radii, indicating a slightly evolved status of the more massive star.  This  can be the reason of the difference in the surface abundance of some elements (different efficiency of diffusion and radiative forces) and in pulsational properties. The agreement with theoretical models required the inclusion of overshooting in the calculations, and the adopted value of $\alpha_{OV}=0.2$ is in agreement with the results of \citet{claret07} for the mass domain considered here. It has to be noticed that the masses of our components are at the lower limit of the Claret sample (whose stars are have all masses above 2 \ms, with the only exception of the secondary of YZ~Cas).
 
 Both the light curve analysis, from the dispersion of the phased light curve, and  that of the radial velocity curves, again from dispersion and from the RME, indicate that the  two dominant frequencies belong to the secondary star. These second  and  first highest amplitude frequencies are identified as  the  fundamental radial mode and a non-radial mode.The most consistent scenario  from the comparison of pulsational analysis with the theoretical modeling indicates as well that the secondary component is at the origin of the high-order g-modes in the $\gamma$~Dor frequency domain, as the driving mechanism is at work only in this component. On the other end this conclusion applies only in the hypothesis of an intrinsic excitation mechanism and  the alternative explanation of  tidal  origin cannot be excluded.  

 Further insight in the study of this interesting system  could be gained by sure identification the pulsation modes, which would increase our knowledge of the internal structure  of the pulsator. That could be achieved by acquiring multicolor photometry and/or    high-S/N high-resolution spectra.  We think that  additional spectroscopy is the best choice: the target is bright enough and the pulsation frequencies not too high  to obtain, even with medium class telescopes (2-4m), high-S/N spectra of high-resolution both in time and wavelength.  Spectroscopic mode identification allows to reach higher $\ell$ values with respect to that by  multi wavelength photometry, which is limited by cancellation effects. Moreover,  spectra of higher S/N than that obtained with the standard follow-up of \kepler targets,   will  provide  a more reliable determination of the atmospheric parameters and hopefully solve the discrepancy on the secondary star temperature and abundances.

 \begin{acknowledgements}
 We thank  Conny Aerts  for  enlightening discussions and critically reading the manuscript, and Daniel Reese, Avi Shporer, Orlagh Creevey,  Kevin Stassun for useful suggestions.
We also express our gratitude to Andrej Pr{\v s}a for making publicly available  PHOEBE and  for his constant  support, to Paul Charbonneau, author of PIKAIA, and to Marc Antoine Dupret for making available to us his computer code MAD. We also thank the unknown referee for constructive criticisms.
This work is partly based on data from the  \kepler mission, which is funded by the NASA Science Mission directorate.  The photometric data were obtained from the Mikulski Archive for Space Telescopes (MAST). STScI is operated by the Association of Universities for Research in Astronomy, Inc., under NASA contract NAS5-26555. Support for MAST for non-HST data is provided by the NASA Office of Space Science via grant NNX13AC07G and by other grants and contracts.
  This research is also partly based on observations obtained with the HERMES spectrograph, which is supported by the Fund for Scientific Research of Flanders (FWO), Belgium, the Research Council of K.U.Leuven, Belgium, the Fonds National Recherches Scientific (FNRS), Belgium, the Royal Observatory of Belgium, the Observatoire de Gen\`{e}ve, Switzerland and the Th\"{u}ringer Landessternwarte Tautenburg, Germany; HERMES is attached to the Mercator Telescope, which is operated on the island of La Palma by the Flemish Community  and is located at the Spanish Observatorio del Roque de los Muchachos of the Instituto de Astrofisica de Canarias.  Additional data were  obtained at the Hobby-Eberly Telescope (HET), a joint project of the University of Texas at Austin, the Pennsylvania State University, Stanford University, Ludwig-Maximilians- Universitat Munchen, and Georg-August-Universitat Gottingen. 
We also made use  of  the VO-KOREL web service, developed at the Astronomical Institute of the Academy of Sciences of the Czech Republic in the framework of the Czech Virtual Observatory (CZVO) by P. Skoda and J. Fuchs using the Fourier disentangling code KOREL  by P. Hadrava.

The authors, finally, acknowledge  generous financial support:
 from the Istituto Nazionale di Astrofisica (INAF)  under  PRIN-2010   {\it Asteroseismology: looking inside the stars with space- and ground-based
observations}  and from the Agenzia Spaziale Italiana (ASI) in the frame of the ESS program  (C.M.  and R.d.S);
 from  BELSPO under contract PRODEX - CoRoT  (J.M.), from NSF under grant AST 1006676 and AST 1126413 (R.D.), from the Turkish Scientific and Research Council (T\"UB\.ITAK 111T270) (KY),
 and from the Center for Exoplanets and Habitable Worlds (R.D.), which is supported by the Pennsylvania State University, the Eberly College of Science, and the Pennsylvania Space Grant Consortium. 
 
\end{acknowledgements}
\bibliographystyle{aa} 
\bibliography{k1e} 
\begin{appendix}
\section{Measured radial velocities}
\begin{table*}[h]
 \caption{Radial velocities (in \kms) of KIC~3858884. 
The columns are BJD 2\,455\,000+ of mid-exposure time, instrument, RV of the primary, RV of the secondary,
and RV of the secondary component cleaned for the short-term variations.
Instruments are B = BOES, E = HET, H = HERMES, and T = TLS spectrograph.}   
\label{rv}      
\centering
\begin{tabular}{ccrrr|ccrrr}
\hline\hline
BJD& source & $RV_1$ & $RV_2$ & $RV'_2$ &
BJD& source & $RV_1$ & $RV_2$ & $RV'_2$ \\
\hline
 503.968710 & B & 101.058 & -69.647 & -71.207 &  751.380682 & T & -16.225 &  49.806 &  48.076\\ 
 509.908470 & B &  -4.498 &  44.847 &  37.951 &  754.353314 & T &  -0.038 &  41.027 &  42.560\\ 
 539.899830 & B & -20.382 &  54.944 &  51.110 &  754.364483 & T &  -3.429 &  39.126 &  39.492\\ 
 541.946970 & B & -18.395 &  46.391 &  51.585 &  754.380861 & T &  -3.786 &  40.539 &  39.212\\ 
 628.303308 & B &  15.562 &  23.515 &  18.222 &  776.461040 & H & -17.718 &  53.411 &  52.308\\
 628.345463 & B &  16.274 &  15.656 &  16.507 &  787.346506 & T &  57.850 & -24.321 & -24.942\\
 629.314475 & B &  22.993 &  -0.129 &  5.260  &  788.485381 & T &  82.243 & -48.560 & -51.161\\
 629.356617 & B &  21.704 &   3.035 &  5.197  &  790.541616 & T & 101.938 & -69.849 & -71.011\\
 633.302954 & B &  94.441 & -61.822 & -63.381 &  791.334012 & T &  79.926 & -44.566 & -47.430\\
 633.345108 & B &  95.566 & -65.888 & -58.958 &  792.326393 & T &  57.915 & -13.379 & -15.017\\
 663.683117 & H &  16.715 &  20.101 &  17.424 &  792.347967 & T &  55.262 & -10.200 & -12.582\\
 664.620059 & H &   1.844 &  26.754 &  28.925 &  793.451780 & T &  20.377 &  15.069 &  10.441\\
 665.732860 & H &  -6.980 &  35.110 &  39.160 &  794.445339 & T &   2.345 &  23.530 &  29.261\\
 666.666483 & H & -12.037 &  38.827 &  45.524 &  800.355096 & T & -19.689 &  52.556 &  51.408\\
 666.719014 & H & -12.116 &  51.810 &  46.882 &  835.485973 & H &  12.871 &  23.537 &  19.594\\
 667.586311 & H & -15.390 &  50.051 &  50.014 & 1049.606437 & H & 107.639 & -77.682 & -73.524\\
 667.727775 & H & -15.555 &  49.056 &  49.725 & 1052.626388 & H &  22.543 &  14.889 &	9.613\\
 681.449628 & T &  23.831 &   4.379 &	3.206 & 1054.636610 & H &  -3.760 &  39.145 &  37.222\\
 682.453035 & T &  39.360 &  -6.949 &  -0.682 & 1056.647034 & H & -15.467 &  54.223 &  50.409\\
 683.564618 & T &  59.074 & -29.658 & -28.887 & 1061.607181 & H & -16.867 &  57.482 &  51.471\\
 692.415476 & T & -10.424 &  45.173 &  44.408 & 1062.689804 & H & -15.963 &  54.937 &  50.092\\
 693.438524 & T & -16.214 &  48.155 &  46.815 & 1064.822359 & E & -11.888 &  40.721 &  40.561\\
 694.150352 & B & -17.183 &  56.479 &  51.729 & 1066.552735 & H &  -2.234 &  40.515 &  35.120\\
 694.506644 & T & -19.691 &  47.490 &  49.977 & 1068.573164 & H &   8.260 &  22.177 &  23.654\\
 695.107980 & B & -19.561 &  56.961 &  51.331 & 1074.838375 & E & 100.575 & -64.819 & -68.291\\
 695.151628 & B & -18.409 &  48.303 &  54.487 & 1076.791640 & E &  79.700 & -49.458 & -49.260\\
 695.193817 & B & -19.763 &  50.567 &  51.356 & 1077.803589 & E &  54.350 & -10.139 & -12.378\\
 696.243405 & B & -19.768 &  48.339 &  52.528 & 1080.772443 & E &  -8.103 &  31.338 &  34.314\\
 697.235861 & B & -18.672 &  51.034 &  51.453 & 1083.761894 & E & -18.470 &  49.227 &  51.194\\
 697.436288 & T & -19.818 &  48.575 &  48.442 & 1086.777471 & E & -18.875 &  54.133 &  51.330\\
 698.239925 & B & -18.455 &  46.693 &  50.055 & 1138.389114 & T & -18.531 &  53.282 &  51.059\\
 700.471873 & T & -12.660 &  50.965 &  46.935 & 1138.530815 & T & -20.323 &  51.781 &  49.585\\
 702.512351 & T &  -4.535 &  32.445 &  37.881 & 1138.586613 & T & -19.753 &  48.648 &  52.199\\
 703.445149 & T &   1.158 &  29.451 &  34.980 & 1139.406973 & T & -17.372 &  43.658 &  49.543\\
 705.428053 & T &   9.932 &  22.247 &  22.467 & 1140.384865 & T & -16.214 &  48.231 &  48.176\\
 716.562951 & H &   3.846 &  26.404 &  29.070 & 1140.450756 & T & -16.666 &  46.626 &  46.029\\
 717.597184 & H &  -5.405 &  40.366 &  39.733 & 1141.585823 & T & -12.740 &  42.465 &  44.432\\
 722.398921 & T & -19.689 &  52.570 &  52.850 & 1143.474649 & T &  -5.791 &  44.063 &  37.751\\
 725.456789 & T & -15.056 &  48.648 &  49.481 & 1144.506855 & T &  -0.162 &  32.431 &  38.587\\
 726.405990 & T & -13.137 &  45.756 &  45.515 & 1148.388880 & T &  21.279 &   0.142 &	5.762\\
 731.416428 & T &  11.112 &  12.745 &  18.390 & 1202.685821 & E &  59.681 & -30.096 & -28.015\\
 734.392309 & T &  39.661 &  -1.017 &  -1.981 &&&&\\
\hline
\end{tabular}
\end{table*}

\section{Frequencies  in the amplitude spectrum of KIC~3858884 from the residuals without eclipses.}
\begin{table*}
 \caption{The first 60 frequencies detected in the amplitude spectrum of KIC~3858884 (from the residuals without eclipses). 
The uncertainties  are derived according to  \citet{kallinger08}. $S$ is the spectral significance
according to \citep{sigspec}. The remark column lists the obvious
overtones/combinations.}   
\label{freqs}      
\centering
\begin{tabular}{lrcccl}
\hline\hline
       &  Frequency (\cd) & Amplitude $(10^{-3})$ &  Phase (2$\pi$)   & $\xi$  & remark \\ 
\hline

$f_{ 1}  $ &    7.2306  $\pm$  0.0001   &   10.15 $\pm$   0.21   &    0.411 $\pm$   0.002   & 2261.2 &      \\
$f_{ 2}  $ &    7.4734  $\pm$  0.0001   &   9.10 $\pm$   0.15   &    0.106 $\pm$   0.001    & 3826.8 &      \\
$f_{ 3}  $ &    9.8376  $\pm$  0.0002   &   1.96 $\pm$   0.07   &    0.190 $\pm$   0.002    &  910.2 &      \\
$f_{ 4}  $ &    7.5125  $\pm$  0.0002   &   1.75 $\pm$   0.06   &    0.646 $\pm$   0.002    &  976.4 &   $f_{ 2}$+f$_{ \rm{orb}}$   \\
$f_{ 5}  $ &    6.7358  $\pm$  0.0002   &   1.55 $\pm$   0.05   &    0.476 $\pm$   0.002    & 1004.1 &      \\
$f_{ 6}  $ &    9.5191  $\pm$  0.0002   &   1.24 $\pm$   0.04   &    0.786 $\pm$   0.003    &  808.5 &      \\
$f_{ 7}  $ &   14.7041  $\pm$  0.0002   &   1.15 $\pm$   0.04   &    0.768 $\pm$   0.002    &  881.9 &   $f_1+f_2$    \\
$f_{ 8}  $ &   11.7257  $\pm$  0.0002   &   1.02 $\pm$   0.04   &    0.575 $\pm$   0.003    &  827.1 &      \\
$f_{ 9}  $ &   14.7253  $\pm$  0.0003   &   0.59 $\pm$   0.03   &    0.330 $\pm$   0.004    &  332.7 &      \\
$f_{ 10}  $ &    7.3628  $\pm$  0.0003   &   0.54 $\pm$   0.03   &    0.264 $\pm$   0.004   &  299.1 &      \\
$f_{ 11}  $ &    7.2424  $\pm$  0.0004   &   0.51 $\pm$   0.03   &    0.351 $\pm$   0.005   &  241.8 &      \\
$f_{ 12}  $ &    7.4621  $\pm$  0.0004   &   0.50 $\pm$   0.03   &    0.988 $\pm$   0.005   &  232.1 &      \\
$f_{ 13}  $ &    0.6971  $\pm$  0.0004   &   0.38 $\pm$   0.03   &    0.721 $\pm$   0.005   &  219.9 &    18  $f_{ \rm{orb}}$, gd    \\
$f_{ 14}  $ &   14.7524  $\pm$  0.0004   &   0.29 $\pm$   0.02   &    0.251 $\pm$   0.005   &  214.5 &     \\
$f_{ 15}  $ &    9.0740  $\pm$  0.0004   &   0.39 $\pm$   0.03   &    0.312 $\pm$   0.005   &  208.6 &      \\
$f_{ 16}  $ &   11.5168  $\pm$  0.0004   &   0.38 $\pm$   0.03   &    0.833 $\pm$   0.005   &  200.6 &      \\
$f_{ 17}  $ &    8.0337  $\pm$  0.0004   &   0.37 $\pm$   0.03   &    0.981 $\pm$   0.005   &  201.8 &      \\
$f_{ 18}  $ &    9.7663  $\pm$  0.0004   &   0.37 $\pm$   0.03   &    0.154 $\pm$   0.005   &  198.9 &      \\
$f_{ 19}  $ &    7.2667  $\pm$  0.0004   &   0.33 $\pm$   0.02   &    0.144 $\pm$   0.005   &  189.1 &     $f_{ 1} + f_{ \rm{orb}}$  \\
$f_{ 20}  $ &    0.0266  $\pm$  0.0005   &   0.70 $\pm$   0.06   &    0.167 $\pm$   0.006   &  162.4 &      \\
$f_{ 21}  $ &   14.2135  $\pm$  0.0005   &   0.37 $\pm$   0.03   &    0.279 $\pm$   0.006   &  160.1 &      \\
$f_{ 22}  $ &    0.2399  $\pm$  0.0005   &   0.15 $\pm$   0.01   &    0.423 $\pm$   0.006   &  154.7 &     $f_{ 2}-f_1$    \\
$f_{ 23}  $ &   10.0012  $\pm$  0.0005   &   0.30 $\pm$   0.02   &    0.070 $\pm$   0.006   &  159.8 &      \\
$f_{ 24}  $ &    9.3034  $\pm$  0.0005   &   0.27 $\pm$   0.02   &    0.844 $\pm$   0.006   &  157.8 &      \\
$f_{ 25}  $ &    7.1830  $\pm$  0.0005   &   0.27 $\pm$   0.02   &    0.103 $\pm$   0.006   &  157.1 &      \\
$f_{ 26}  $ &   17.0682  $\pm$  0.0005   &   0.26 $\pm$   0.02   &    0.541 $\pm$   0.006   &  148.4 &     $f_{ 1}+f_3$     \\
$f_{ 27}  $ &   14.4613  $\pm$  0.0005   &   0.27 $\pm$   0.02   &    0.300 $\pm$   0.006   &  131.0 &    2 $f_{ 1}$     \\
$f_{ 28}  $ &   13.8340  $\pm$  0.0005   &   0.23 $\pm$   0.02   &    0.539 $\pm$   0.007   &  126.7 &      \\
$f_{ 29}  $ &   14.9470  $\pm$  0.0005   &   0.23 $\pm$   0.02   &    0.317 $\pm$   0.007   &  128.5 &  2 $f_{ 2}$    \\
$f_{ 30}  $ &    9.2801  $\pm$  0.0005   &   0.39 $\pm$   0.03   &    0.725 $\pm$   0.006   &  131.9 &      \\
$f_{ 31}  $ &   17.1049  $\pm$  0.0005   &   0.23 $\pm$   0.02   &    0.726 $\pm$   0.007   &  128.7 &      \\
$f_{ 32}  $ &   10.3419  $\pm$  0.0005   &   0.23 $\pm$   0.02   &    0.728 $\pm$   0.007   &  128.1 &      \\
$f_{ 33}  $ &   16.9926  $\pm$  0.0005   &   0.22 $\pm$   0.02   &    0.491 $\pm$   0.007   &  122.1 &      \\
$f_{ 34}  $ &    6.9632  $\pm$  0.0005   &   0.22 $\pm$   0.02   &    0.933 $\pm$   0.007   &  123.9 &      \\
$f_{ 35}  $ &   14.1544  $\pm$  0.0005   &   0.21 $\pm$   0.02   &    0.494 $\pm$   0.007   &  118.6 &      \\
$f_{ 36}  $ &   11.7733  $\pm$  0.0005   &   0.21 $\pm$   0.02   &    0.584 $\pm$   0.007   &  119.7 &      \\
$f_{ 37}  $ &    0.1544  $\pm$  0.0005   &   0.21 $\pm$   0.02   &    0.238 $\pm$   0.007   &  119.4 &   4   $f_{ \rm{orb}}$   \\
$f_{ 38}  $ &    9.1235  $\pm$  0.0005   &   0.20 $\pm$   0.02   &    0.708 $\pm$   0.007   &  114.4 &      \\
$f_{ 39}  $ &    0.0099  $\pm$  0.0005   &   0.19 $\pm$   0.02   &    0.353 $\pm$   0.007   &  115.9 &      \\
$f_{ 40}  $ &   14.2097  $\pm$  0.0005   &   0.29 $\pm$   0.03   &    0.596 $\pm$   0.007   &  116.4 &      \\
$f_{ 41}  $ &    0.9020  $\pm$  0.0005   &   0.20 $\pm$   0.02   &    0.990 $\pm$   0.007   &  115.3 &   gd   \\
$f_{ 42}  $ &   14.7485  $\pm$  0.0006   &   0.33 $\pm$   0.03   &    0.332 $\pm$   0.007   &  109.9 &  $f_{ 1}+ f_{ 4}$     \\
$f_{ 43}  $ &    0.0416  $\pm$  0.0006   &   0.33 $\pm$   0.03   &    0.793 $\pm$   0.007   &  102.4 &      \\
$f_{ 44}  $ &   14.9860  $\pm$  0.0006   &   0.17 $\pm$   0.02   &    0.305 $\pm$   0.007   &   98.5 &   2$f_{ 2} + f_{ \rm{orb}}$      \\
$f_{ 45}  $ &    7.2038  $\pm$  0.0006   &   0.17 $\pm$   0.02   &    0.479 $\pm$   0.008   &   85.6 &   $f_{ 1} - f_{ \rm{orb}}$   \\
$f_{ 46}  $ &   11.2943  $\pm$  0.0006   &   0.16 $\pm$   0.02   &    0.337 $\pm$   0.008   &   86.9 & $f_{ 2} + f_{ 5}$        \\
$f_{ 47}  $ &    0.5114  $\pm$  0.0006   &   0.17 $\pm$   0.02   &    0.987 $\pm$   0.008   &   87.4 &   gd   \\
$f_{ 48}  $ &    0.5058  $\pm$  0.0006   &   0.14 $\pm$   0.01   &    0.543 $\pm$   0.008   &   90.3 &      \\
$f_{ 49}  $ &    6.8510  $\pm$  0.0006   &   0.15 $\pm$   0.02   &    0.688 $\pm$   0.008   &   90.9 &  $ f_{ 5} + 3 f_{{\rm orb}}$    \\
$f_{ 50}  $ &    7.5006  $\pm$  0.0006   &   0.18 $\pm$   0.02   &    0.600 $\pm$   0.008   &   85.8 &      \\
$f_{ 51}  $ &    6.2352  $\pm$  0.0006   &   0.15 $\pm$   0.02   &    0.489 $\pm$   0.008   &   85.8 &      \\
$f_{ 52}  $ &    3.4906  $\pm$  0.0006   &   0.14 $\pm$   0.02   &    0.653 $\pm$   0.008   &   83.5 &      \\
$f_{ 53}  $ &    0.0625  $\pm$  0.0007   &   0.13 $\pm$   0.01   &    0.050 $\pm$   0.008   &   78.4 &      \\
$f_{ 54}  $ &   14.7414  $\pm$  0.0007   &   0.18 $\pm$   0.02   &    0.352 $\pm$   0.009   &   76.0 &      $f_{ 1} +f_{ 2}+ f_{ \rm{orb}}$\\
$f_{ 55}  $ &    1.0539  $\pm$  0.0007   &   0.13 $\pm$   0.02   &    0.078 $\pm$   0.008   &   77.0 &   gd  \\
$f_{ 56}  $ &    2.3796  $\pm$  0.0007   &   0.14 $\pm$   0.02   &    0.927 $\pm$   0.008   &   76.2 &  \\
$f_{ 57}  $ &    0.2185  $\pm$  0.0007   &   0.39 $\pm$   0.04   &    0.166 $\pm$   0.008   &   76.5 &  \\
$f_{ 58}  $ &    2.3641  $\pm$  0.0007   &   0.14 $\pm$   0.02   &    0.620 $\pm$   0.008   &   77.6 & \\
$f_{ 59}  $ &    7.4232  $\pm$  0.0007   &   0.13 $\pm$   0.01   &    0.938 $\pm$   0.008   &   78.3 &  \\
$f_{ 60}  $ &    7.2797  $\pm$  0.0007   &   0.17 $\pm$   0.02   &    0.743 $\pm$   0.008   &   77.6 &  \\

\hline
\end{tabular}
\end{table*}

\end{appendix}

 \end{document}